\documentclass[graybox, secnum]{svmult}

\usepackage{mathptmx}       % selects Times Roman as basic font
\usepackage{helvet}         % selects Helvetica as sans-serif font
\usepackage{courier}        % selects Courier as typewriter font
\usepackage{type1cm}        % activate if the above 3 fonts are
\usepackage{makeidx}         % allows index generation
\usepackage{graphicx}        % standard LaTeX graphics tool
\usepackage{multicol}        % used for the two-column index
\usepackage[bottom]{footmisc}% places footnotes at page bottom
\usepackage{hyperref}        %for hyperlinks
\usepackage{soul}            % for high-lighting of text
\usepackage[normalem]{ulem}  % for \sout{} Pedro
\usepackage{aas_macros_prc} %includes abbreviations for AAS journals
\hypersetup{colorlinks=true,urlcolor=blue}
\usepackage[square,numbers]{natbib}
\makeindex             % used for the subject index
\usepackage{amsmath,amssymb}%Pedro

\newcommand{\doi}[1]{\textcolor{black}{#1}}%Pedro
\newcommand{\comments}[1]{} %usage: \comments{}

\begin{document}

\title*{Black hole-galaxy co-evolution and the role of feedback}

\author{Pedro R. Capelo\thanks{Center for Theoretical Astrophysics and Cosmology, Institute for Computational Science, University of Zurich, Winterthurerstrasse 190, CH-8057 Z{\"u}rich, Switzerland; \email{pcapelo@physik.uzh.ch} (corresponding author)},
Chiara Feruglio\thanks{INAF Osservatorio Astronomico di Trieste, Via G.B. Tiepolo 11, IT-34143 Trieste, Italy; Institute for Fundamental Physics of the Universe, Via Beirut 2, IT-34014 Trieste, Italy; \email{chiara.feruglio@inaf.it} (corresponding author)},
Ryan C. Hickox\thanks{Department of Physics and Astronomy, Dartmouth College, Hanover, NH 03755, USA; \email{Ryan.C.Hickox@dartmouth.edu} (corresponding author)},
Francesco Tombesi\thanks{Department of Physics, University of Rome ``Tor Vergata'', Via della Ricerca Scientifica 1, IT-00133 Rome, Italy; INAF Osservatorio Astronomico di Roma, Via di Frascati 33, IT-00040 Monteporzio Catone, Rome, Italy; Department of Astronomy, University of Maryland, College Park, MD 20742, USA; NASA Goddard Space Flight Center, Code 662, Greenbelt, MD 20771, USA; \email{francesco.tombesi@roma2.infn.it} (corresponding author)}}
\maketitle

\setcounter{footnote}{0}

\abstract{Active galactic nuclei (AGN) are accreting supermassive black holes co-evolving with their host galaxies through a complex interplay of feeding and feedback. In this chapter, we first discuss AGN fuelling in galaxies, both in interacting and isolated systems, focusing on the role that instabilities have on the angular momentum budget of the gas. We then review observations and models of feedback through AGN-driven winds from nuclear, sub-pc scales out to galactic and circumgalactic medium scales. We continue with an overview of surveys and statistical properties of the AGN population, before concluding with a discussion on the prospects for the future facilities, focusing in particular on \emph{Athena}.}


\clearpage

\section{AGN fuelling}\label{sec:AGN_fuelling}%Pedro R. Capelo

Black holes (BHs) are amongst the most efficient energy converters in the Universe: when mass is accreted on to a BH, a significant fraction of its rest energy is released, most of which as radiation. This can be quantified with the (dimensionless) radiative efficiency, $\epsilon_{\rm r}$, defined via

\begin{equation}
    L_{\rm BH} = \epsilon_{\rm r}\,\dot{M}_{\rm BH,accr}\,c^2,
    \label{eq:LBH}
\end{equation}

\noindent where $L_{\rm BH}$ is the radiated bolometric luminosity (energy per unit time), $\dot{M}_{\rm BH,accr}$ is the accretion rate (accreted mass per unit time),\footnote{The corresponding BH growth rate is defined as $\dot{M}_{\rm BH,growth} = (1 - \epsilon_{\rm m})\dot{M}_{\rm BH,accr}$, where $\epsilon_{\rm m}$ is the fraction of rest mass energy released by accretion in any form. In this section, we will assume for simplicity $\epsilon_{\rm r} = \epsilon_{\rm m}$, but note that $\epsilon_{\rm r}$ can be $< \epsilon_{\rm m}$, due to, e.g. winds/jets.} and $c$ is the speed of light in vacuum.

When detecting\footnote{What is usually detected is a small fraction of $L_{\rm BH}$, e.g. between 0.01 and 0.1~$L_{\rm BH}$ in the hard X-rays (2-10~keV), or $\sim$0.2~$L_{\rm BH}$ in the $\lambda$ 4400~{\AA} $B$-band (e.g. \cite{Duras_et_al_2020}), not accounting for obscuration.} or simulating accreting supermassive BHs (SMBHs), a typical value of $\epsilon_{\rm r} = 0.1$ is usually adopted, which is consistent with the So\l{}tan argument (\cite{Soltan_1982}): by comparing the SMBH mass density at redshift $z = 0$ to the integrated luminosity of active galactic nuclei (AGN) over time, and assuming that most of the SMBH mass growth is indeed due to gas accretion, one obtains a rough match between the two quantities when adopting a radiative efficiency of $\sim$10 per cent (e.g. \cite{Marconi_et_al_2004}). This efficiency is extremely large, especially when compared to that of chemical reactions ($\sim$10$^{-10}$), or even to the nuclear fusion of hydrogen ($\sim$0.7 per cent). In individual cases, the radiative efficiency can be even higher, depending on both the nature of the BH and on the geometry of the accreting matter. This matter is usually assumed to take the form of an accretion disc, wherein gas loses angular momentum, moves inwards, and radiates. Assuming, for example a Novikov--Thorne (\cite{Novikov_Throne_1973}) disc around a maximally spinning (prograde) Kerr BH (\cite{Kerr_1963}), $\epsilon_{\rm r}$ can surpass 40 per cent.\footnote{In reality, the maximum allowed spin for a BH surrounded by an accretion disc is slightly lower than the theoretical limit of 1, taking the maximum radiative efficiency down to $\sim$0.3 (\cite{Thorne_1974}).}

Despite the very high values of the radiative efficiency, the enormous amounts of radiated energy per unit time observed in many AGN still imply a formidable supply of accreted gas. For a given radiated luminosity, the accretion rate required can be written as

\begin{equation}
    \dot{M}_{\rm BH,accr} \simeq 2 \times 10^{-3} \left(\frac{L_{\rm BH}}{10^{43} {\rm erg\, s}^{-1}}\right)\left(\frac{0.1}{\epsilon_{\rm r}}\right){\rm M}_{\odot}\,{\rm yr}^{-1}
    \label{eq:Mdot}
\end{equation}

\noindent or, alternatively, as

\begin{equation}
    \dot{M}_{\rm BH,accr} \simeq 2 \,f_{\rm Edd} \left(\frac{M_{\rm BH}}{10^8 \,{\rm M}_\odot}\right) \left(\frac{0.1}{\epsilon_{\rm r}}\right) {\rm M}_{\odot}\,{\rm yr}^{-1},
    \label{eq:Mdot2}
\end{equation}

\noindent where $f_{\rm Edd} \equiv L_{\rm BH}/L_{\rm Edd}$ and

\begin{equation}
    L_{\rm Edd} = \frac{4 \pi G M_{\rm BH} m_{\rm p} c}{\sigma_{\rm T}}
    \label{eq:LEdd}
\end{equation}

\noindent is the Eddington luminosity (\cite{Eddington_1916}), with $G$ being the gravitational constant, $m_{\rm p}$ the proton mass, and $\sigma_{\rm T}$ the Thomson cross-section. $L_{\rm Edd}$ is computed assuming spherical symmetry and imposing a balance between the inward gravitational force and the outward radiation force on an electron--proton pair. The corresponding Eddington accretion rate, $\dot{M}_{\rm Edd}$, is simply given by imposing $f_{\rm Edd} = 1$ in Equation~\eqref{eq:Mdot2}, usually assuming $\epsilon_{\rm r} = 0.1$.

In order to sustain an AGN radiating a bolometric luminosity of $10^{46-48}$~erg~s$^{-1}$ (i.e. $2.6 \times 10^{12-14}$~L$_\odot$; or the Eddington luminosity of an SMBH of $\sim$8$ \times 10^{7-9}$~M$_\odot$), the BH requires $\sim$2--200~M$_\odot$~yr$^{-1}$ of fuel (adopting $\epsilon_{\rm r} = 0.1$). If we now assume a typical AGN lifetime of $\sim$10$^8$~yr (either continuous or as the sum of shorter bursts; e.g. \cite{Marconi_et_al_2004,Schawinski_et_al_2015}), $\sim$2$\times 10^{8-10}$~M$_{\odot}$ of gas are required: this is quite a difficult task, considering that such an amount is comparable to the total gas reservoir in many galaxies. Though these numbers may seem extreme, they fall in the same ballpark of what is required to grow SMBHs with masses of the order of $10^9$~M$_\odot$ found at $z > 7.5$ (e.g. \cite{Wang_et_al_2021}), when the Universe was a few hundred Myr old (see Chapter~4 in this Volume).

A challenge much more daunting than having enough gas is delivering this fuel to the SMBH. The gas present in the BH's host must travel from galactic distances (i.e. of the order of a few or few tens of kpc) down to the BH's innermost stable circular orbit (ISCO). The ISCO is thought to be roughly the inner edge of the accretion disc and, in the case of a maximally spinning (prograde) SMBH of $10^8$~M$_\odot$, has a radius of 1~AU. In order to travel this far down to the centre, the galactic gas, which is usually in the form of a large-scale disc (similar in size to the stellar disc), needs to lose copious angular momentum (e.g. \cite{Jogee_2006,Alexander_Hickox_2012,Capelo_2019}).

As an illustrative example, we quantify the required loss by appraising the specific angular momentum of the gas in spiral galaxies at three characteristic radii: the BH's ISCO radius, the BH's radius of influence, and typical large-scale distances.

Spiral galaxies exhibit a correlation between the SMBH's mass and the maximum disc rotational velocity, $v_{\rm rot}$ (\cite{Davis_et_al_2019}):

\begin{equation}
    \log_{10}\left(\frac{M_{\rm BH}}{{\rm M}_\odot}\right) = (10.62 \pm 1.37) \log_{10}\left(\frac{v_{\rm rot}}{210 {\rm \,km \,s}^{-1}}\right) + (7.22 \pm 0.09).
    \label{eq:M-v}
\end{equation}

Thus, in the case of an SMBH of $10^8$~M$_\odot$, $v_{\rm rot} \sim 2.5 \times 10^2$~km~s$^{-1}$ and the specific angular momentum of gas at a distance $r = 10$~kpc from the centre, assuming a flat rotational velocity curve, is $l_{\rm 10 \,kpc} = r \,v_{\rm rot} \sim 2.5 \times 10^3$~kpc~km~s$^{-1}$.

We can compute the same quantity at the radius of influence of the BH (within which the BH's gravitational potential is dominant), commonly defined as

\begin{equation}
    r_{\rm BH,inf} = \frac{G M_{\rm BH}}{\sigma^2},
    \label{eq:r_BH-inf}
\end{equation}

\noindent where $\sigma$ is the stellar velocity dispersion, which we assume to be constant in the inner galactic region. The velocity dispersion can be inferred from the so-called $M_{\rm BH}$--$\sigma$ relation (e.g. \cite{Ferrarese_Merritt_2000,Gebhardt_et_al_2000,Kormendy_Ho_2013,McConnell_Ma_2013}),

\begin{equation}
    \log_{10}\left(\frac{M_{\rm BH}}{{\rm M}_\odot}\right) = (3.19 \pm 1.24) \log_{10}\left(\frac{\sigma}{141 {\rm \,km \,s}^{-1}}\right) + (7.31 \pm 4.51),
    \label{eq:M-sigma}
\end{equation}

\noindent where we combined the $M_{\rm BH}$--$v_{\rm rot}$ and $v_{\rm rot}$--$\sigma$ relations from \cite{Davis_et_al_2019} for consistency with Equation~\eqref{eq:M-v}.\footnote{The quoted slope in \cite{Davis_et_al_2019} ($M_{\rm BH} \propto \sigma^{3.21 \pm 1.19}$) is recovered within the accuracy of their $v_{\rm rot}$--$\sigma$ relation.} Assuming that the central region can be described by a singular isothermal sphere, the gas then has a circular velocity of $\sqrt{2}\sigma$. Considering the same SMBH of $10^8$~M$_\odot$ as above, we obtain $r_{\rm BH,inf} \sim 8$~pc and $\sigma \sim 2.3 \times 10^2$~km~s$^{-1}$, and the specific angular momentum at the BH's radius of influence is $l_{r_{\rm BH,inf}} = \sqrt{2} r_{\rm BH,inf} \sigma \sim 2.6$~kpc~km~s$^{-1}$.

Moving further down, we can calculate the specific angular momentum of gas at the BH's ISCO radius (below which the gas spirals inwards): in the case of a maximally spinning (prograde) BH, this is simply $l_{\rm ISCO} = c r_{\rm S}/\sqrt{3}$ (\cite{Ruffini_Wheeler_1971}; where $r_{\rm S} = 2 G M_{\rm BH}/c^2$ is the Schwarzschild \cite{Schwarzschild_1916} radius), which, for a $10^8$~M$_\odot$ SMBH, yields $1.7 \times 10^{-3}$~kpc~km~s$^{-1}$.\footnote{This quantity is 3 and 11/3 times larger (at a given BH mass) for non-spinning and maximally spinning (retrograde) BHs, respectively  (\cite{Ruffini_Wheeler_1971}). Note that the Newtonian approximation made in \cite{Capelo_2019} is only a factor of $2/\sqrt{3}$ smaller.}

\begin{figure}[t!]
\centering
\vspace{-0.35cm}
\includegraphics[scale=0.75]{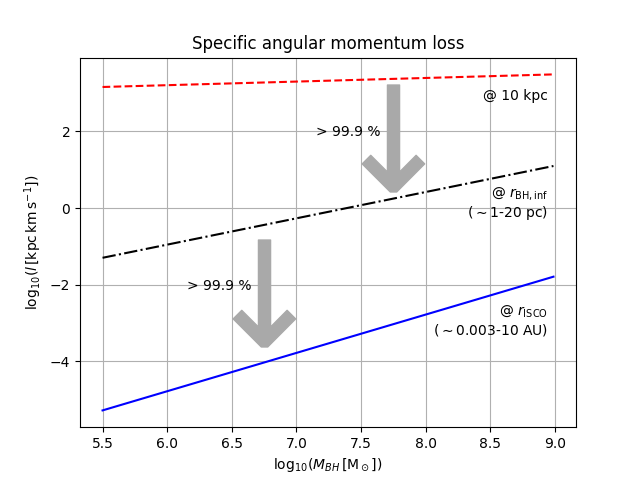}
\vspace{-0.5cm}
\caption{Specific angular momentum of the gas in a spiral galaxy containing a central bulge and a (maximally spinning, prograde) SMBH, as a function of the SMBH mass: at 10~kpc from the centre (dashed, red line), at the BH's influence radius (dash-dotted, black line), and at the BH's ISCO radius (solid, blue line). For clarity, we used only the mean values of the coefficients in Equations~\eqref{eq:M-v} and \eqref{eq:M-sigma}. {\it The specific angular momentum of the gas travelling from galactic scales down to the BH's ISCO must decrease by more than 99.999 per cent.}}
\vspace{-0.5cm}
\label{fig:capelo_ang_mom_loss}
\end{figure}

These numbers do not strongly depend on the mass of the SMBH, since $l_{\rm 10 \,kpc} \propto M_{\rm BH}^{0.09}$, $l_{r_{\rm BH,inf}} \propto M_{\rm BH}^{0.69}$, and $l_{\rm ISCO} \propto M_{\rm BH}$, as shown in Figure~\ref{fig:capelo_ang_mom_loss} (where the SMBH mass range is consistent with \cite{Davis_et_al_2019}).

This means that, for a wide range of SMBH masses, {\it the specific angular momentum of the gas has to decrease by roughly three orders of magnitude when moving from galactic scales down to the BH's sphere of influence and by another three orders of magnitude from $r_{\rm BH,inf}$ to the BH's ISCO, for a total loss of more than 99.999 per cent.}

The study of gas transport encompasses all scales from extremely large (e.g. cosmological gas inflows) to extremely small (e.g. accretion discs). In this section, we assume that galaxies already have their gas, i.e. we do not treat cosmological inflows nor cluster-scale phenomena, such as cooling flows or ram-pressure stripping (for the same reason, we focus on disc galaxies, since early-type galaxies are similar in nature to galaxy clusters). Here, we limit ourselves to the topic of gas inflows on galactic-to-nuclear scales, i.e. from kpc- to pc-scales.

In the next two sub-sections, we discuss how gas can reach the nuclear regions of galaxies in both interacting and isolated systems.

\subsection{Interacting galaxies}\label{sec:Interacting_galaxies}%Pedro R. Capelo

\vspace{-0.4cm}
Galactic interactions have been recognised as one of the obvious drivers for the inflow of gas within galaxies since the first numerical studies of mergers that incorporated a collisional component (see, e.g. \cite{Icke_1985,Sorensen_1985,Noguchi_Ishibashi_1986}): the rapidly varying gravitational potentials and the extreme pressure gradients, both a natural consequence of mergers, offer clear physical mechanisms for gas transport within the interacting systems.

Tidal torques (due to gravitational forces) caused by interactions between galaxies can be extremely effective at reducing the angular momentum of the gas: each companion causes the formation of large non-axisymmetric structures, both gaseous and stellar (such as bars), in the other galaxy. These structures do not exactly overlap, due to the different nature of gas and stars: the collisionless stellar structure trails behind the collisional gaseous configuration and torques the gas, inducing an inflow (see, e.g. \cite{Noguchi_1988,Barnes_Hernquist_1991,Barnes_Hernquist_1996,Hopkins_et_al_2009,Hopkins_Quataert_2010}).\footnote{Other gravitational torques (e.g. directly from the companion or from the halo) are also present but are not as effective (see, e.g. \cite{Barnes_Hernquist_1996}). Note that interactions can also destroy an existing bar, albeit for a limited amount of time (e.g. \cite{Zana_et_al_2018b}), thus hindering the gas inflow.} The efficiency of this mechanism decreases as the gas nears the centre, since large-scale bars tend to be perturbed (and destroyed) near the inner Lindblad resonance\footnote{The inner Lindblad resonance occurs when the epicyclic frequency of a star is twice the difference between the circular frequency and the pattern speed of the bar.} radius (e.g. \cite{Kormendy_2013}). The inflow, however, can continue owing to a succession of ``non-axisymmetric features all the way down'', including nuclear bars (\cite{Hopkins_Quataert_2010}; see also \cite{Shlosman_et_al_1989,Shlosman_et_al_1990,Jogee_2006}).

Working concurrently are hydrodynamical torques, which arise from pressure gradient forces. Though considered sub-dominant in early studies (due to relatively poor numerical resolution; e.g. \cite{Barnes_Hernquist_1991}), the strength of this mechanism has been recently shown to be comparable to that of tidal torques, especially at pericentre passages and closer to the centre (\cite{Prieto_et_al_2021}; but see \cite{Blumenthal_Barnes_2018}). A sub-class is ram pressure, when large-scale ($\sim$kpc) shocks produce a strong deceleration of the gas of one of the galaxies (or both), which decouples the dynamics of the collisional component from that of stars \cite{Barnes_2002,Capelo_Dotti_2017,Blumenthal_Barnes_2018}: these ram-pressure shocks can cause the gas to lose angular momentum and flow inwards, even if the tidal-triggered non-axisymmetric structures present orbital resonances which tend to stop the gas transport (\cite{Capelo_Dotti_2017}).

Shocks can also produce clumps from Jeans-unstable filaments, which then travel to the centre via dynamical friction (\cite{Blumenthal_Barnes_2018}), possibly resulting in a more intermittent delivery of gas to the central regions of the galaxies. The resultant AGN activity, however, is contingent on the lifetime of these gas clumps, which depends on the efficiency and time-scale of star formation and on the strength of stellar feedback\footnote{Stellar feedback is the injection of energy, momentum, and metals into the interstellar medium (ISM) during the life and death (as supernovae) of stars.} (e.g. \cite{Hopkins_Quataert_2010,Hopkins_et_al_2013}): if gas in these clumps undergoes star formation, then only stars would be delivered to the central galactic region.\footnote{Depending on the mass of the central BH, it is however possible to observe another type of (short-lived) BH activity, namely tidal disruption events (TDEs; bright electromagnetic flares occurring when a star is torn apart by the tidal forces in the vicinity of a massive BH ($\lesssim 10^8$~M$_{\odot}$); see, e.g. \cite{Gezari_et_al_2021}), whose rates are indeed enhanced during mergers (\cite{Pfister_et_al_2019}).}

\begin{figure}[t!]
\centering
\includegraphics[scale=0.25]{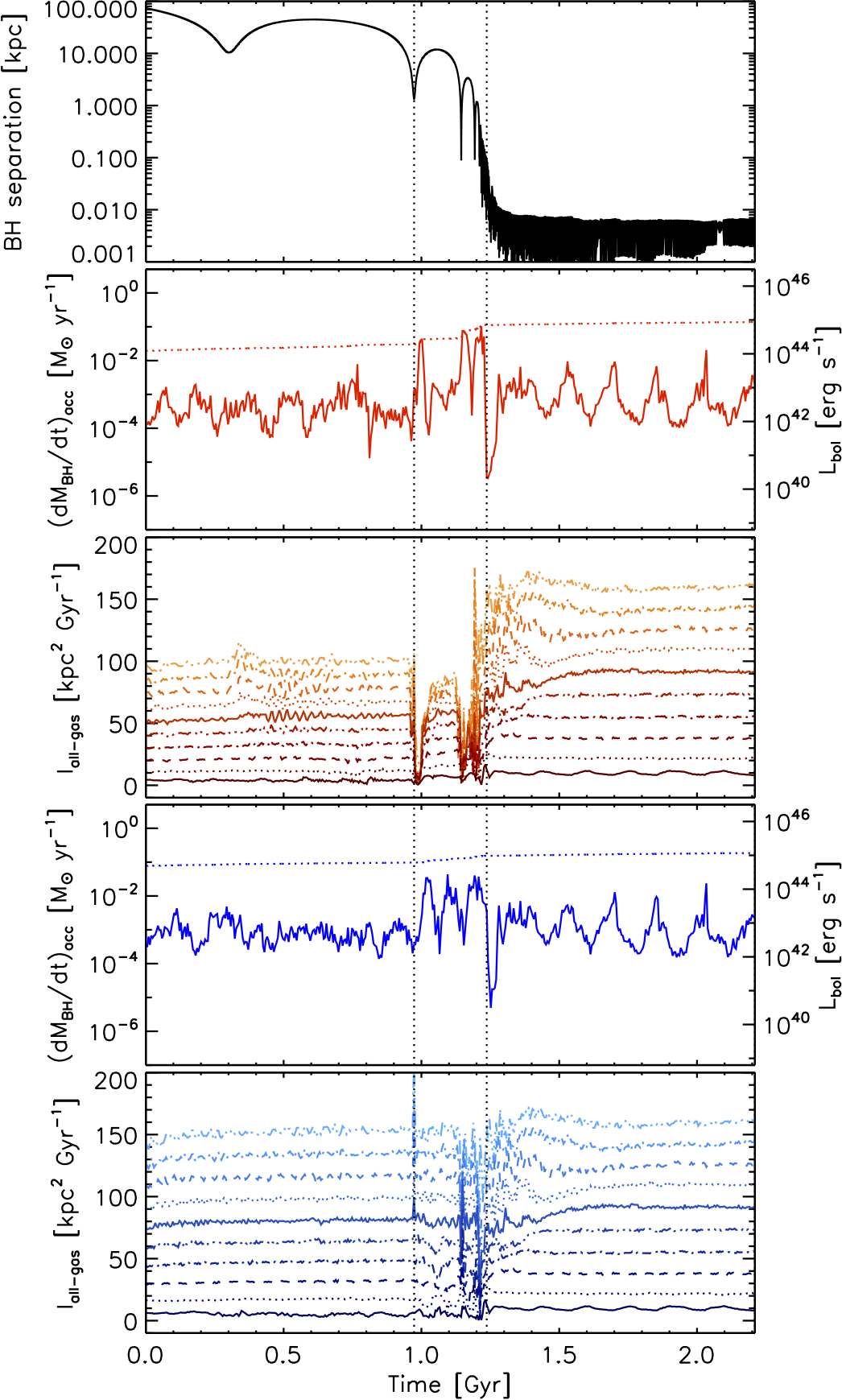}
\caption{Time evolution of the gas specific angular momentum and BH accretion rate of two interacting galaxies in a simulated 1:4 merger presented in \cite{Capelo_et_al_2015}. The first panel shows the BH separation. The second and fourth panels show the BH accretion rate (and related bolometric luminosity; see Equation~\ref{eq:LBH}; solid line) and Eddington luminosity (see Equation~\ref{eq:LEdd}; dotted line) of the secondary and primary BH, respectively. The third and fifth panels show the specific angular momentum of the gas of the secondary and primary galaxy, respectively, in ten concentric central shells of 100~pc width (from dark to light colour as the radius of the shell increases). In all panels, the two vertical, dotted lines show the beginning and end of the proper merger stage. {\it The specific angular momentum of the gas and the BH accretion rate are anticorrelated, at all encounter stages and for both galaxies, although the secondary galaxy is more strongly affected.} Figure adapted from figures~1 and 2 of \cite{Capelo_et_al_2015}.}
\label{fig:capelo_merger}
\end{figure}

As an illustrative case, we discuss one of the simulated mergers presented in \cite{Capelo_et_al_2015}, where the initial mass ratio of the interacting galaxies (and their BHs) is 1:4 and the internal angular momenta of the discs are aligned with the total orbital angular momentum. In Figure~\ref{fig:capelo_merger}, we show the time evolution of the main properties discussed in this section, namely the specific angular momentum of the gas of the merging galaxies (third and fifth panel) and the accretion rate of the two BHs (second and fourth panel), along with the BH separation (first panel). The accretion on to the BHs is stochastic and relatively low both before and after the proper merger (the time between the two vertical lines), when the gas is not strongly affected by the interaction. However, during the merger phase (a period of $\sim$$2 \times 10$$^8$~yr following the second pericentric passage), the gas loses most of its angular momentum. This occurs especially in the secondary galaxy, which, being much lighter, is more strongly affected by the interaction. At the same time of these central angular momentum troughs, we observe high peaks of BH accretion rate.

\clearpage%to force the figure in the right place

This distinct correlation, which exists also before and after the proper merger phase (and for both galaxies, although less visible in the primary), is also evident when comparing BH accretion rate\footnote{In \cite{Capelo_et_al_2015}, as in the majority of numerical works on the same topic, BH accretion is computed using a version of the Bondi--Hoyle--Lyttleton (hereafter Bondi) accretion recipe (\cite{Hoyle_Lyttleton_1939,Bondi_Hoyle_1944,Bondi_1952}), in which $\dot{M}_{\rm Bondi} \propto M_{\rm BH}^2$ and capped at $\alpha \dot{M}_{\rm Edd}$ (see Equations~\ref{eq:Mdot2} and \ref{eq:LEdd}; with $\alpha$ of order unity), whereas BH feedback is modelled as a spherically symmetric release of energy per unit time (to the nearest gas particles) proportional to Equation~\eqref{eq:LBH} (see \cite{Bellovary_et_al_2010}). Similar results were also obtained, using the same setup but with a different numerical code, in \cite{Gabor_et_al_2016}.} and central star formation rate (SFR; \cite{Volonteri_et_al_2015a,Volonteri_et_al_2015b}), hinting at the fact that, although in competition for the same fuel (cold, dense gas), the two phenomena can co-exist, at least during galactic interactions, when inflows are substantial (see also \cite{Hobbs_et_al_2011,Alexander_Hickox_2012,Hickox_et_al_2014}). These inflows are caused by a combination of tidal torques and ram-pressure shocks (\cite{Capelo_Dotti_2017}), which are also responsible for angular momentum flips, when the angular momentum of the entire disc changes by 180$^\circ$ for a few $10^8$~Myr, during the most extreme phases of the encounter (see also \cite{VanWassenhove_et_al_2014}).

Note that what matters is not the magnitude of the specific angular momentum, but its derivative with respect to time: in other words, it is the gas that loses angular momentum (and not necessarily that with low angular momentum) that gets accreted on to the BH \cite{Capelo_et_al_2015}.

The time-scales and efficiency of the fuelling mechanisms depend on several properties of the merging galaxies, and the results just shown can vary significantly: these properties include galaxy morphology (e.g. an elliptical galaxy contains relatively little cold gas; \cite{VanWassenhove_et_al_2012}), bulge-to-disc stellar mass ratio (e.g. a larger bulge tends to stabilise the disc to perturbations; \cite{Blecha_et_al_2013}), gas fraction (e.g. a lower gas fraction leads to the formation of a stronger bar; \cite{Blumenthal_Barnes_2018}), and encounter geometry (e.g. a coplanar merger tends to maximise inflows; \cite{Capelo_et_al_2015}).

The most important quantity, however, is the secondary-to-primary mass ratio, $q_{\rm BH} \equiv M_{\rm BH,2}/M_{\rm BH,1}$ (assuming here, for simplicity, that the initial $q_{\rm BH}$ is equal to the initial mass ratio between the hosts). On the one hand, the primary galaxy is more affected in major mergers (i.e. those with high mass ratio), since only in those the companion is massive enough to significantly affect the larger galaxy. On the other hand, the secondary galaxy tends to always respond strongly to the interaction (regardless of the mass ratio). This has noticeable effects on the strength of the gas inflows and on the ensuing BH accretion rate (see also, e.g. \cite{Cox_et_al_2008}, for a similar effect on star formation), with minor BH mergers tending to become less minor (${\rm d}q_{\rm BH}/{\rm d}t > 0$) and major BH mergers tending to become less major (${\rm d}q_{\rm BH}/{\rm d}t < 0$; \cite{Capelo_et_al_2015}). For the same reason, the mass ratio is crucial for determining the detectability of dual AGN activity, with major mergers showing a longer-lasting activity than minor mergers (\cite{VanWassenhove_et_al_2012,Blecha_et_al_2013,Capelo_et_al_2017,Blecha_et_al_2018,DeRosa_et_al_2019,Volonteri_et_al_2021}).

We note that the same mechanisms and inflows discussed in this section are not only relevant for AGN fuelling, but are important also for the formation itself of SMBHs (e.g. via the gravitational collapse of an unstable, massive central gas disc fuelled by merger-driven inflows; \cite{Mayer_et_al_2010,Mayer_et_al_2015}) and for the SMBH pairing time-scales (e.g. aided by merger-fed massive nuclear discs and/or stellar nuclei; \cite{Escala_et_al_2005,Dotti_et_al_2006,Dotti_et_al_2007,Mayer_et_al_2007,VanWassenhove_et_al_2014,SouzaLima_et_al_2020}).

As discussed in this section, there is a clear consensus on galactic encounters being able to disturb the baryonic components of the interacting systems, both the stellar constituent (e.g. \cite{Holmberg_1941,Toomre_Toomre_1972,White_1978,Tamfal_et_al_2018}) and its gaseous counterpart (e.g. \cite{Barnes_Hernquist_1992,Mihos_Hernquist_1996,Cox_et_al_2008}), thus potentially leading to gas inflows that are able to feed AGN activity. Such a causal connection has been amply documented in numerical studies (e.g. \cite{DiMatteo_et_al_2005,Capelo_et_al_2015}) and further indirectly bolstered by an extensive observational literature showing a clear increase in the AGN fraction in close pairs of galaxies (e.g. \cite{Ellison_et_al_2011,Satyapal_et_al_2014}). This, however, does not necessarily mean that galaxy mergers are the dominant triggering mechanism for all AGN, and indeed there have been many studies both in favour of (e.g. \cite{Ellison_et_al_2019}) and against (e.g. \cite{Cisternas_et_al_2011}) this hypothesis (see, e.g. \cite{Storchi-Bergmann_Schnorr-Muller_2019} for a recent review): these opposing results are likely due to a combination of selection effects (e.g. \cite{Ellison_et_al_2019}), AGN variability (see, e.g. \cite{Schawinski_et_al_2015,Comerford_et_al_2017}), obscuration (e.g. \cite{Hopkins_et_al_2006}), and dependence on luminosity (e.g. \cite{Treister_et_al_2012}, but see \cite{Hewlett_et_al_2017}) and redshift (e.g. \cite{Marian_et_al_2019,Marian_et_al_2020}).

Regardless of the outcome of this dispute, it is a fact that a significant fraction of AGN activity has been observed in non-interacting systems: in the next section, we discuss AGN fuelling in isolated galaxies.

\subsection{Isolated galaxies}\label{sec:Isolated_galaxies}%Pedro R. Capelo

Angular momentum loss can occur also in isolated systems, due to a variety of disc instabilities. Local disc galaxies are usually subject to weak non-axisymmetric instabilities, whereas high-redshift galaxies, with their high gas fractions, can undergo violent clump instabilities.

In the previous section, we saw that non-axisymmetric structures can arise as a consequence of interactions. These same structures (and bars in particular), however, can also form in isolation (and, in fact, the origin of bars is still debated; see, e.g. \cite{Zana_et_al_2018a}), when the gravitational potential of the galaxy becomes prone to bar instability. The galactic gas is then torqued and flows to the centre. The bar-induced mechanism is believed to work mostly in the local Universe, since smooth gas discs are usually found at relatively low redshift (although recent numerical and observational studies have shown the existence of discs and even barred discs at very high redshift; e.g. \cite{Bortolas_et_al_2020,Rizzo_et_al_2020,Kretschmer_et_al_2022,Tamfal_et_2022}). As in the case of mergers and AGN, the connection between BH fuelling and large-scale bars is also debated (see, e.g. \cite{Jogee_2006,Storchi-Bergmann_Schnorr-Muller_2019}), as there are more observed barred systems than AGN: this may be partially due to the fact that large-scale bars are effective at transporting gas inwards only down to $\sim$100~pc, or also because major gas inflows occur mostly during the formation of the bar, as observed in \cite{Fanali_et_al_2015}, or simply a consequence of AGN variability.

\begin{figure}[t!]
\centering
\includegraphics[scale=0.175]{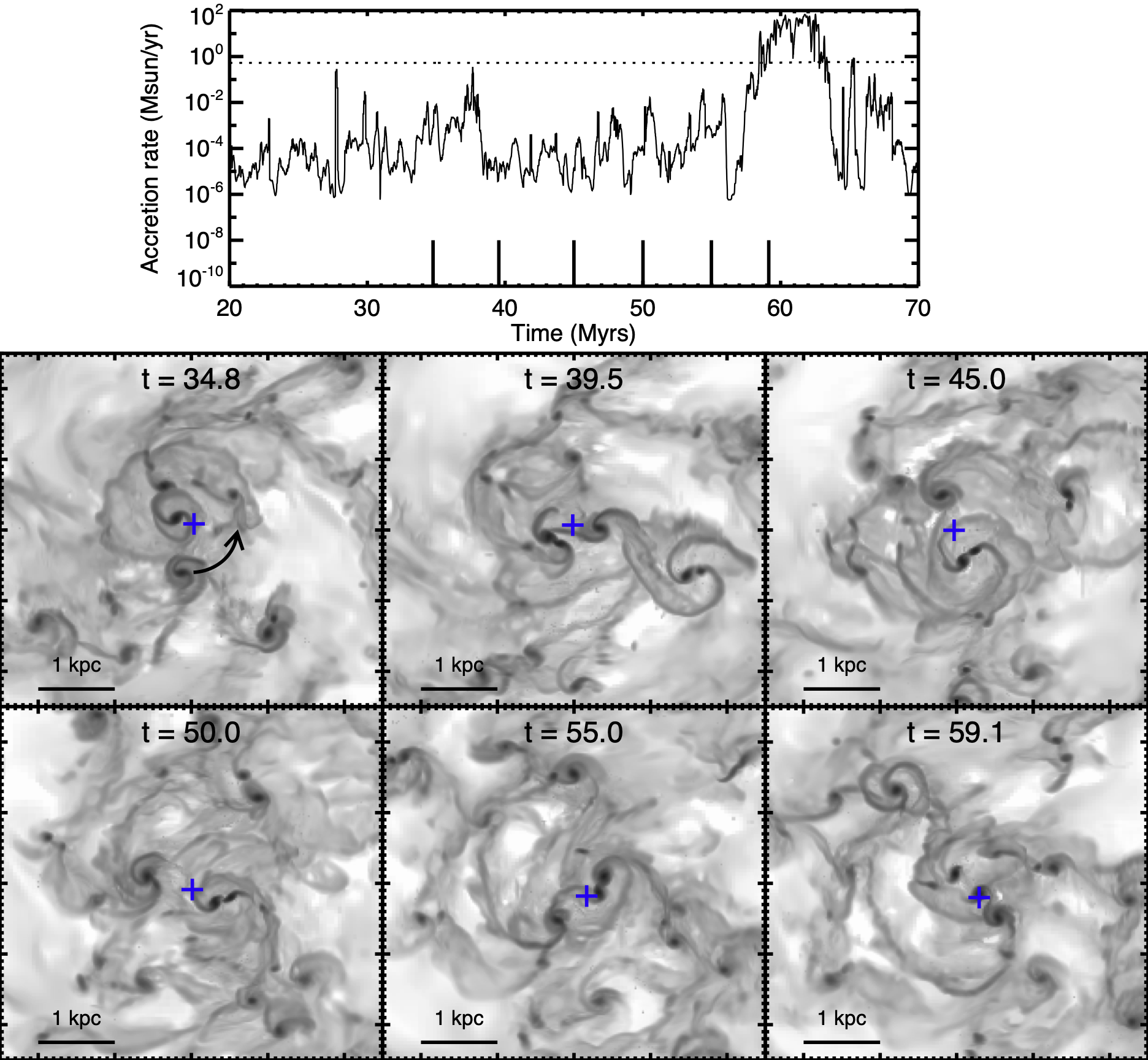}
%\vspace{5pt}
\caption{Top panel: time evolution of the Bondi accretion rate on to the central SMBH of a simulated clumpy galaxy presented in \cite{Gabor_Bournaud_2013}. The dotted line shows the Eddington limit ($\dot{M}_{\rm Edd}$): the actual accretion rate in the code is the minimum between $\dot{M}_{\rm Bondi}$ and $\dot{M}_{\rm Edd}$. Bottom panels: face-on, projected mass-weighted gas density maps of the isolated galaxy, at six different times (marked by the vertical lines in the top panel). The blue cross shows the location of the central SMBH. {\it Collisions between dense clouds and the SMBH} (like the one shown at time 59.1~Myr) {\it can trigger episodes of Eddington-limited growth.} Figure from \cite{Gabor_Bournaud_2013}.}
\label{fig:gabor_clumps}
\end{figure}

Since the association between bars and AGN fuelling has been discussed in the previous section (although in the context of galactic interactions), we will here focus on another mechanism, the feeding of BHs by violent clump instabilities.

Giant gas clumps are believed to form especially in high-redshift, massive gaseous discs: because of their high surface density (and corresponding low Toomre parameter: $Q = \kappa c_{\rm s}/(\pi G \Sigma)$, where $\kappa$ is the epicyclic frequency, and $c_{\rm s}$ and $\Sigma$ are the sound speed and surface density of the gas, respectively; \cite{Toomre_1964}), these discs are prone to gravitational instabilities and to subsequent fragmentation. These clumps can then feed the central region of the galaxy (and the BH itself) in two complementary ways: migration and inflows (see, e.g. \cite{Bournaud_2016}). Clumps can migrate inwards via dynamical friction against the background baryonic material (see also \cite{Blumenthal_Barnes_2018}, in the context of mergers) and by exchanging angular momentum with over-dense regions (gaseous spiral arms) induced by the clumps themselves. These same over-dense regions, together with the gas clumps, exchange angular momentum with the rest of the disc, inducing an inflow of gas (e.g. \cite{Bournaud_et_al_2011}). The masses and lifetimes\footnote{Also within isolated clumpy galaxies, as it was in the case of galaxy mergers, if gas clumps do not reach the centre before undergoing star formation, it is still possible to detect BH activity in the form of TDEs (see \cite{Pestoni_et_al_2021}).} of these clumps are crucial for assessing their effects on the BH accretion rate and are still the topic of much ongoing investigation (see, e.g. \cite{Tamburello_et_al_2015}).

This global inflow of gas and its connection to BH activity are illustrated in Figure~\ref{fig:gabor_clumps}, which shows the temporal evolution of the Bondi accretion rate on to an SMBH at the centre of a simulated $z \sim 2$ gas-rich galaxy, along with gas density maps of the system at a few representative times (\cite{Gabor_Bournaud_2013}). The massive disc is clearly unstable to fragmentation into giant gas clumps, which stochastically collide with the central BH (see also \cite{Hopkins_Hernquist_2006,DeGraf_et_al_2017}), producing episodes of high levels of accretion, such as that at $\sim$60~Myr (see the bottom-right density map). Similar collisions and increases in AGN activity may occur also for non-central BHs (the result of, e.g. galaxy mergers), as seen, e.g. in simulations of isolated galaxies (\cite{Tamburello_et_al_2017}) and their central regions (\cite{SouzaLima_et_al_2017}).

It is natural then to assume a link between clumpy galaxies and AGN activity, although its strength is still debated (see, e.g. \cite{Bournaud_et_al_2012,Trump_et_al_2014}), possibly due to obscuration (\cite{Bournaud_et_al_2011}) or small samples. Since the fraction of clumpy galaxies at high redshift may be significant ($>$0.5 at $1 \lesssim z \lesssim 2.5$ for star-forming galaxies with typical UV luminosities; \cite{Shibuya_et_al_2016}), more work (both numerical and observational) should be devoted to understanding this connection.

\section{AGN feedback}%Chiara & Francesco

There are several indications of relations between SMBHs and their host galaxies. For instance, the $M_{\rm BH}$--$\sigma$ relation shows that the bigger the mass of the SMBH, the higher is the velocity dispersion of stars in the galaxy bulge (e.g. \cite{Kormendy_Ho_2013}). Moreover, large-scale simulations of galaxy evolution show that the high-mass end of the galaxy stellar mass function is over-predicted and some phenomena should be responsible for the quenching of the star formation. One promising possibly being AGN feedback (e.g. \cite{bower2012}). Therefore, a fundamental question is, how do SMBHs affect galaxy evolution? 

The typical radius of an SMBH is a billion times smaller than that of a galaxy. Moreover, the typical mass of an SMBH is less than 1 per cent of the stellar bulge mass (e.g. \cite{magorrian1998}). However, it is important to note that the total gravitational potential energy of SMBHs ($\propto M_{\rm BH} c^2$) is comparable to or higher than the binding energy of the entire galaxy bulge ($\propto M_* \sigma_{*}^{2}$)! In this calculation we can use a typical mass of a galaxy bulge of $M_*\sim 10^{11}$~M$_{\odot}$, a stellar velocity dispersion of $\sigma_*\sim 100$~km~s$^{-1}$, and an SMBH mass of $M_{\rm BH}\sim 10^7$~M$_{\odot}$. 

Therefore, if there is a way to tap into just a fraction this energy, SMBHs may be able to exert a strong impact on the host galaxy. Indeed, the conversion of gravitational energy into radiation or jets and winds can have a profound impact on the AGN host galaxy. Here, we focus on the AGN-driven winds, which are thought to be the agents of the so-called ``quasar-mode feedback'' (e.g. \cite{fabian2012,king2015}). Differently from the radio-mode feedback, which involves a small fraction of the AGN population, AGN winds are ubiquitous in AGN, and therefore the quasar-mode feedback can potentially affect the bulk of the AGN population.

Accretion and ejection mechanisms on to SMBHs are intimately related, because the accretion disc (i.e. feeding, described in the previous section), naturally gives rise to winds (i.e. feedback), and in turn outflows of matter affect both the SMBH growth and the environment. In this section we give an overview of AGN-driven winds at different scales, from nuclear, sub-pc scales out to galactic and circumgalactic medium (CGM) scales, and involving different gas phases, from highly ionized to cold atomic and molecular gas. 

\subsection{Warm absorbers}%Francesco

\begin{figure}[h!]
\centering
\includegraphics[scale=1]{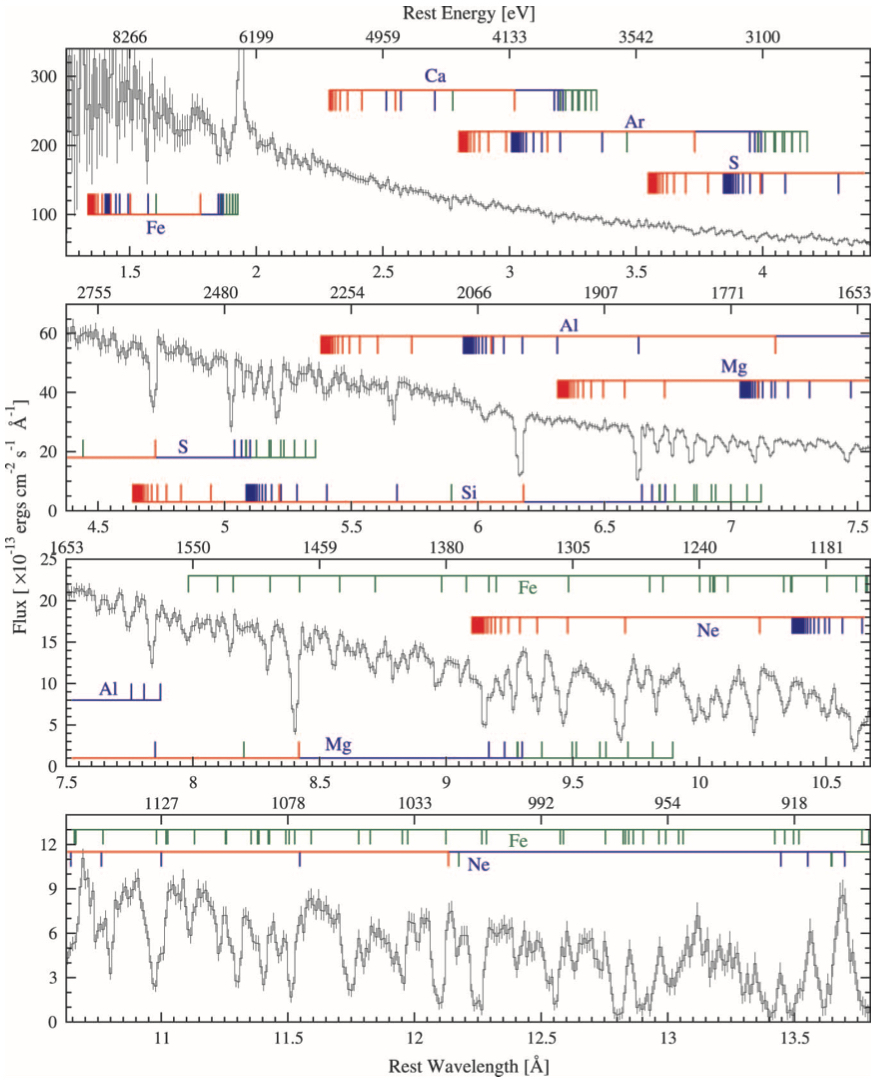}
\caption{Broad-band Chandra high-energy transmission grating (HETG) spectrum of NGC 3783 showing X-ray absorption and emission lines due to the WA (figure from \cite{kaspi2002}).}
\label{fig:kaspi2002}
\end{figure}

The first evidence for photo-ionized absorption in the X-ray spectrum of an AGN was reported by \cite{halpern1984} comparing two spectra of the quasar MR 2251-178 taken in 1979 and 1980 with the Einstein satellite. The data showed a change in the soft X-ray band, which was interpreted as a ``warm'' photo-ionized, rather than ``cold'' , cloud causing a large change in the observed column density of $N_{\rm H}\simeq6\times 10^{22}$~cm$^{-2}$ along the line of sight (LOS) to the quasar. This type of absorber was referred to as a ``warm absorber'' (WA) because the temperature was estimated to be $T\sim 10^5$~K, higher than cold neutral gas but much cooler than that required by a thermal plasma. 

The advent of high-energy resolution grating spectrometers onboard Chandra and XMM-Newton since the year 2000 provided a revolution in these studies, showing that the absorption is composed of a number of lines and edges from different elements at different ionization states (e.g. \cite{kaspi2002}). Importantly, the energies of these lines are found to be systematically blue-shifted compared to the expected values, indicating that the material is likely in the form of a wind outflowing from the central regions of AGN with velocities in the range of $v_{\rm out} \simeq 100$--1000~km~s$^{-1}$ (e.g. \cite{mckernan2007,kaastra2014}). 

To understand the physics and impact of these outflows, we need to have an estimate of their physical and dynamical parameters. From spectral fitting of absorption lines/edges in X-rays with photo-ionization codes, such as Cloudy (\cite{ferland2017}) or XSTAR (\cite{kallman2004}), one can obtain the ionization parameter ($\xi$), the equivalent hydrogen column density ($N_{\rm H}$), the outflow velocity ($v_{\rm out}$), and an estimate of the velocity width, or turbulence ($\sigma$). The turbulence velocity is not necessarily physically linked to actual turbulence in the gas, but it is used to parameterize random motion along the line of sight in excess to the thermal broadening. The ionization parameter is defined as $\xi = L_{\rm ion}/n_{\rm e} r^2$ (\cite{tarter1969}), where $L_{\rm ion}$ is the ionizing luminosity of the source integrated from 13.6 eV to 13.6 keV, $n_{\rm e}$ is the electron density of the plasma, and $r$ is the distance of the cloud from the ionizing source.

The absorption lines arise from different ionic states of astronomically abundant elements such as O, Ne, Si, S, Fe, and others. Overall, such WAs are detected in more than half of local Seyfert galaxies and have ionization and column densities in the range log$\xi \simeq 0$--3~erg~s$^{-1}$~cm and $N_{\rm H} \simeq 10^{20}$--$10^{22}$ cm$^{-2}$, respectively (e.g. \cite{crenshaw2012}). Often, WAs are modelled with multiple absorption components with different parameters. Figure~\ref{fig:kaspi2002} clearly shows the soft X-ray WA in the high-energy resolution Chandra HETG spectrum of the local Seyfert galaxy NGC~3783 (\cite{kaspi2002}).

Given their intermediate ionization levels, possible relations between WAs and winds detected in the optical/UV have been reported, suggesting that we may see the same material both in soft X-rays and UV (e.g. \cite{crenshaw2012}). Furthermore, WAs have been detected also in some radio-galaxies, suggesting that the production of winds and jets is not mutually exclusive, depending on the accretion state of the AGN (e.g. \cite{torresi2012,tombesi2016}).

\subsection{Ultra-fast outflows}%Francesco

\begin{figure}[h]
\centering
\includegraphics[scale=0.5]{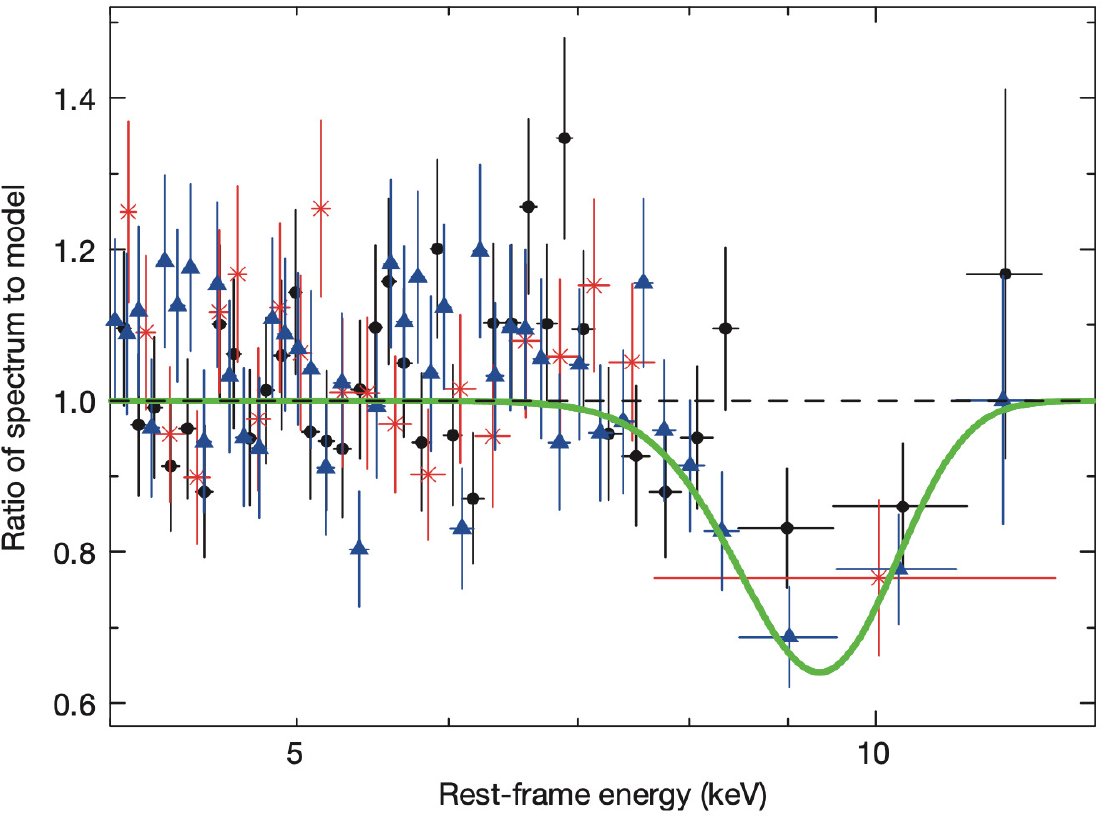}
\caption{Ratio between the Suzaku spectrum and an absorbed power-law continuum model in the $E=4$--12~keV energy range, clearly showing the broad UFO absorption feature at $E\simeq 9$~keV in IRAS~F11119$+$3257. The black filled circles, red asterisks, and blue triangles represent the Suzaku XIS0, XIS1, and XIS3 data, respectively. Figure from \cite{tombesi2015}.}
\label{fig:tombesi2015}
\end{figure}

Thanks to the high effective area and high-energy coverage of the spectrometers onboard XMM-Newton, Suzaku and NuSTAR, evidence for a more extreme type of outflow compared to the WA emerged in the X-ray spectra of AGN. These outflows were detected preferentially at rest-frame energies higher than 7~keV through blue-shifted absorption lines from highly ionized iron (e.g. \cite{chartas2002,pounds2003}). Extensive analyses on larger samples of AGN, mostly local ($z<0.1$) X-ray bright (4--10~keV flux $>10^{-12}$ erg~s$^{-1}$~cm$^{-2}$) Seyferts, showed that these so-called ultra-fast outflows (UFOs) are present in at least 40 per cent of the sources (\cite{tombesi2010,gofford2013}). The UFOs are highly ionized, with ionization parameter log$\xi \simeq 3$--6~erg~s$^{-1}$~cm,  and have high column densities in the range $N_{\rm H} \simeq 10^{22}$--$10^{24}$~cm$^{-2}$. 

Most importantly, their implied outflow velocities are often mildly relativistic, in ranging from $\simeq 10^4$~km~s$^{-1}$ up to significant fractions of the speed of light, $\simeq$0.3--$0.5c$ (\cite{tombesi2011,gofford2013}). This is a fundamental parameter that indicates that the UFO are most likely ejected close to the central SMBH and that they can carry a large kinetic power into the environment of the AGN host galaxy (e.g. \cite{tombesi2012,gofford2015}). Figure~\ref{fig:tombesi2015} clearly shows the UFO absorption feature in the ratio between the Suzaku X-ray spectrum and the absorbed power-law continuum model in the $E=4$--12~keV energy range of IRAS~F11119$+$3257 (\cite{tombesi2015}). 

By comparing samples of radio-loud and radio-quiet AGN, it has been found that the UFO detection fractions in both these cases are similar (50$\pm$20 per cent), implying that the presence of relativistic jets does not preclude the existence of these winds (e.g. \cite{tombesi2010b,tombesi2014}). The UFO detection fraction is a useful parameter which suggests that their covering fraction of the X-ray source and/or that their duty cycle is $\sim$50 per cent. 

Additional very important insights into the geometry and kinematics of UFOs may come from the detection of emission lines associated with the same material seen in absorption along the line of sight. Indeed, some sources show the so-called ``P-Cygni'' profiles, i.e. profiles in the Fe K band, which allow one to estimate the opening angle and covering fraction of the associated outflow. For instance, \cite{nardini2015} reported the combined XMM-Newton and NuSTAR spectrum of the luminous quasar PDS~456 (at $z=0.184$), an archetype for radio-quiet AGN with powerful disc winds, which suggests a P-Cygni profile due to a large opening angle UFO with a covering fraction of $\simeq$80 per cent.

High-redshift quasars are an important laboratory for studying the AGN phenomenon, as these are the periods (redshift $z \sim 1$--3) when the quasar luminosity function peaked (see Chapter 1 of this Volume). It is also commonly believed that the SMBH ``negative'' feedback may have produced quasar downsizing, and hence studying feedback through outflows in the quasars at these redshifts is important for understanding the luminosity function evolution. Early studies detected high-velocity Fe XXVI absorption features in the lensed quasar APM~08279$+$5255 at a redshift of $z = 3.91$, which shows also broad absorption line (BAL) wind features in the UV spectrum (e.g. \cite{chartas2002}). High-redshift studies to date remain relatively scarce due to the limited signal-to-noise and mostly rely on lensed sources. Next-generation instruments, particularly \emph{Athena}, may shed more light on these interesting objects thanks to the very high sensitivity.

Recently, \cite{chartas2021} presented results from a comprehensive study of UFOs in a larger sample of high-redshift quasars. The derived outflow velocities, ranging between $\simeq 0.1c$ and $\simeq 0.6c$, seem significantly higher than those found in low-redshift AGN and they seem to be launched from very close to the SMBH ($r < 100 r_{\rm g}$, where $r_{\rm g} = GM_{\rm BH}/c^2$). Special relativistic corrections become significant for such high velocities (e.g. \cite{luminari2020}). Consequently,  their kinetic power is also found to be rather large, typically higher than half of their bolometric luminosities. This suggests that they may provide efficient feedback to influence the evolution of their host galaxies and a magnetic driving (i.e. magnetic pressure and/or magneto-centrifugal forces, see also Section~2.3) may be a significant contributor to their acceleration (e.g. \cite{fukumura2010}). 

Apart from the ``classical'' UFOs detected in the energy range $E=6$--9~keV due to Fe K-shell absorption features, there has been an increasing number of detections of high-velocity outflows using high-resolution grating spectra in the soft X-rays ($E=0.3$--2~keV) (e.g. \cite{longinotti2015,gupta2015,reeves2020}). The UFOs detected in the soft X-ray spectra show velocities up to $\sim$$0.2c$, but with a much lower column density ($N_{\rm H} \simeq 10^{20}$--$10^{22}$ cm$^{-2}$) and ionization parameter (log$\xi \simeq 0$--3~erg~s$^{-1}$~cm) compared to the UFOs detected from Fe K-shell lines. These outflows could represent lower-ionization clumps of gas within UFOs or the outcome of an interaction of the UFO with the host galaxy environment. 

Finally, high-ionization UFOs have often been found to be variable even on short time-scales ($\sim$hrs) and sometimes to respond to variations of the AGN continuum. These studies are fundamental to infer the possible density, location and dynamics of the outflows, which are then used to derive their energetics (e.g. \cite{parker2017}).

\subsection{Scaling relations for X-ray winds}%Francesco

\begin{figure}[ht]
\centering
\includegraphics[scale=1.3]{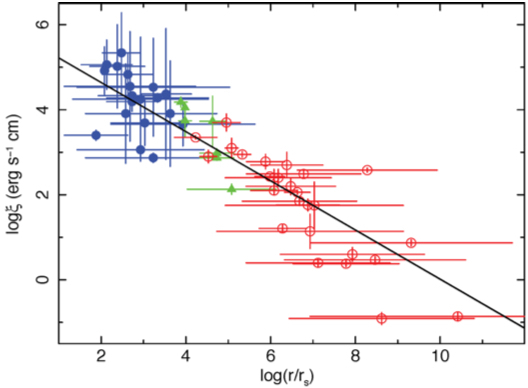}
\includegraphics[scale=0.6]{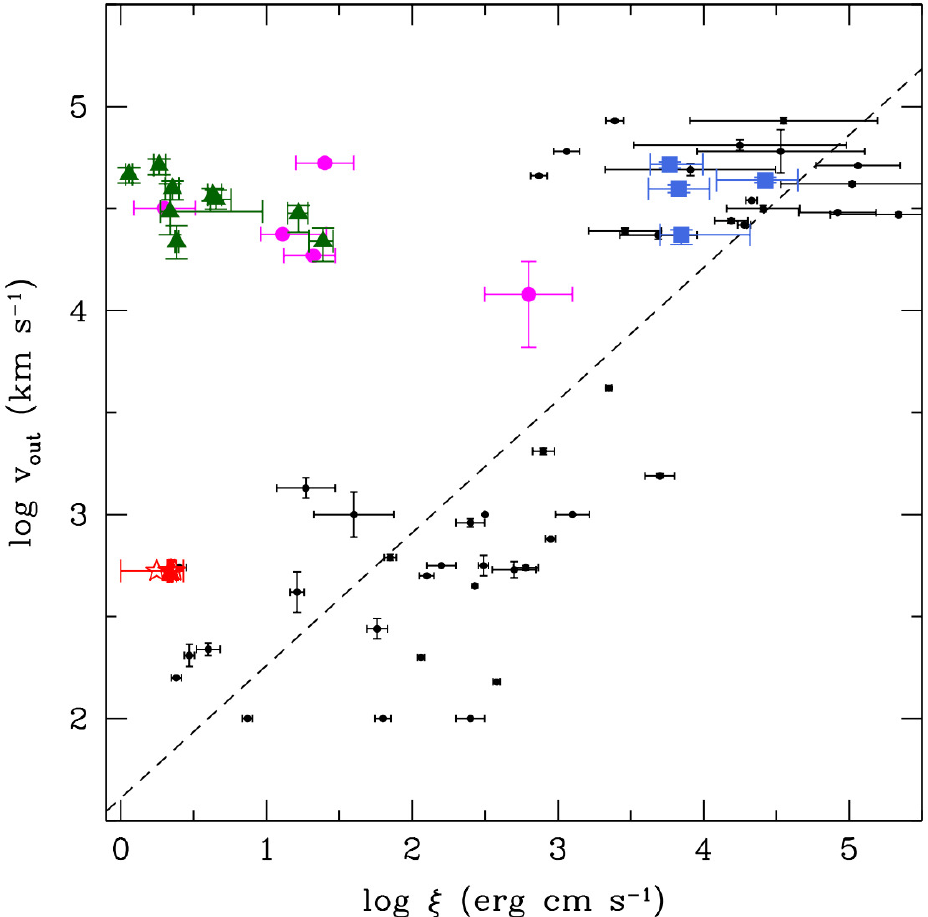}
\caption{Left-hand panel: comparison between the distance log($r/r_{\rm S}$) (in units of the Schwarzschild radius) and the ionization parameter log$\xi$ for the WAs (red open circles), non-UFOs (green filled triangles), and UFOs (blue filled circles) in the local Seyfert sample of \cite{tombesi2013}. The solid line indicates the best-fitting linear regression curve. Assuming the typical Seyfert SMBH mass of $M_{\rm BH} \simeq 10^7$~M$_{\odot}$, the distance scale can be easily converted from the Schwarzschild radii to parsec considering that 1~pc $\simeq$ $10^6$~$r_{\rm S}$. Right-hand panel: velocity versus ionization parameter plot for WA and UFO detected in a sample of local AGN (black points). The dashed line shows a linear fit in log-log space (from \cite{tombesi2013}).  The WA (red stars), UFO (blue squares), and E-UFO (green triangle) detected in the X-ray spectrum of PG~1114$+$445 are also shown. The magenta points represent additional soft X-ray UFOs from the literature (figure from \cite{serafinelli2019}).}
\label{fig:serafinelli2019}
\end{figure}

The large dynamic range obtained when considering WAs and UFOs together, spanning several orders of magnitude in ionization (log$\xi$$\sim$0--6 erg~s$^{-1}$~cm), column ($N_{\rm H}$$\sim$$10^{20}$--$10^{24}$ cm$^{-2}$), velocity ($v_{\rm out}$$\sim 10^2$--$10^5$ km~s$^{-1}$) and distance (sub-pc to kpc) from the SMBH is an important tool to investigate overall relations and trends among them (e.g. \cite{tombesi2013,laha2021}). As it can be seen in the left-hand panel of Figure~\ref{fig:serafinelli2019}, the absorbers populate the whole parameter space, with the WAs and the Fe K UFOs lying at the two ends of the distribution (upper left and bottom right, respectively). This evidence suggests that these X-ray absorbers, often considered independently, may represent parts of a large-scale stratified outflow observed at different locations from the SMBH. In particular, the closer is the absorber to the central SMBH, the higher are the values of the wind ionization, column, and velocity, and consequently its power. 

The UFOs are likely launched from the inner accretion disc and the WAs at larger distances, such as the outer disc and/or torus. The detailed acceleration mechanisms of disc winds in AGN are still matter of intense research. However, they can be broadly classified in thermal, radiation and magnetohydrodynamic (MHD), depending on their main driving force. Given the intense radiation field in most AGN, radiation pressure itself can be a very effective way to drive a disc wind up to speeds of $\simeq 0.1c$ (e.g. \cite{elvis2000,proga2000}). Another promising possibility to explain the origin of disc winds in AGN (and possibly also X-ray binaries) is offered by MHD models. In this case, the wind is accelerated by the centrifugal force of the magnetic field lines anchored on the disc and the magnetic pressure (e.g. \cite{blandford1982,fukumura2010}). The wind is ejected from different regions of the disc. This causes a stratification, with increasing column density, ionization and velocity closer to the central SMBH, which may resemble the trends observed in the left-hand and right-hand panels of Figure~\ref{fig:serafinelli2019}.

An alternative origin for, at least some, WA components could derive from the interaction of the UFO with the larger-scale galactic environment. Supporting this idea, it has been conjectured that the WAs may be generated via a Compton-cooled shocked wind, when a UFO launched very near to the SMBH loses most of its mechanical energy after shocking against the ISM (e.g. \cite{pounds2013}). In this regard, \cite{serafinelli2019} reported the spectral analysis of 12 XMM-Newton observations of the quasar PG~1114$+$445, aimed at studying the complex outflowing nature of its absorbers. The analysis revealed the presence of three absorbing structures: a typical soft X-ray WA, an Fe K UFO with $v  \simeq 0.15c$, and an additional absorber in the soft X-rays ($E < 2$~keV) with velocity comparable to that of the UFO ($v_{\rm out} \simeq 0.12 c$), but with ionization (log$\xi \simeq 0.5$~erg~s$^{-1}$~cm) and column density ($N_{\rm H} \simeq 21.5$ cm$^{-2}$) comparable with those of the WA. The ionization, velocity, and variability of the three absorbers indicate an origin in a multiphase and multiscale outflow, consistent with entrainment of the clumpy ISM by an inner UFO, producing an entrained ultra-fast outflow (E-UFO). The right-hand panel of Figure~\ref{fig:serafinelli2019} shows the parameters of the WA, E-UFO and UFO in PG~1114$+$445 compared to those of X-ray winds detected in other local AGN. 
  
In order to assess the possible impact of outflows on AGN feedback, it is imperative to estimate their energetics.  Assuming the simplified case of a quasi-spherical, radial wind, we can express the mass outflow rate using the following formula: $\dot{M}_{\rm out} = 1.23 m_{\rm p} n r^2 C_{\rm v} C_{\rm g} v_{\rm out}$~g~s$^{-1}$. Here, the factor 1.23 accounts for the cosmic abundance of elements compared to hydrogen and $m_{\rm p}$ is the proton mass. The main parameters to be estimated are the outflow velocity $v_{\rm out}$, directly measured in the spectrum, the density $n$, the location $r$, the volume filling factor $C_{\rm v}$, and the global covering factor of the wind $C_{\rm g}$. Often, $C_{\rm v}$ is approximated with unity and $C_{\rm g}$ with the fraction of detected outflows in a population of sources or, when present, using the equivalent width of wind emission lines. The density and location are very difficult to estimate and often only lower or upper limits can be placed combining the estimated ionization parameter and column density or assuming the location at which the outflow velocity is equal to the escape velocity. Knowing the mass outflow rate, we can then estimate the outflow momentum rate or force as $\dot{P}_{\rm out} = M_{\rm out} v_{\rm out}$ dyne, and the instantaneous mechanical power or kinetic luminosity $\dot{E}_{\rm out} = \frac{1}{2} \dot{M}_{\rm out} v_{\rm out}^2$ erg~s$^{-1}$. 

Extensive studies on UFOs (\cite{tombesi2012,tombesi2013,gofford2015,chartas2021}) show that they are likely launched from the accretion disc at distances of $\sim$10--10$^{3}$ $r_{\rm g}$ from the SMBH, with a mass outflow rate comparable to the accretion rate up to $\sim$1~M$_{\odot}$~yr$^{-1}$. The kinetic power is also found to be systematically larger than a few per cent of the AGN bolometric luminosity, indicating that UFO are indeed capable to drive AGN feedback in the host galaxy. Moreover, correlations with the continuum are found: the velocity of UFOs scales with the AGN bolometric luminosity $v_{\rm out} \propto L_{\rm bol}^{\alpha}$ with $\alpha \simeq 0.4$, and the momentum rate of the UFOs is proportional to the AGN radiation momentum rate, $\dot{M}_{\rm out}v_{\rm out} \simeq L_{\rm bol}/c$ (\cite{tombesi2012,tombesi2013,gofford2015}). These evidences suggest a link between the accretion of matter on to SMBHs and the launching of UFOs, possibly through radiation pressure and/or magnetic forces. 

On the other hand, WAs seems to the located at scales larger than a pc from the AGN and do not show clear relations with the AGN luminosity. Their kinetic power is likely not sufficient for driving significant AGN feedback, but they may transport a significant amount matter in the host galaxy environment (\cite{crenshaw2012,tombesi2013}).

\subsection{Winds on galactic scales}%Chiara

Besides the highly-ionized semi-relativistic winds seen in the previous section, there are winds that involve cooler gas phases, that are the warm ionized ISM, the neutral-atomic and the cold molecular ISM. These cooler winds are widespread in AGN host galaxies at all redshifts, and constitute the preferred feedback carriers in radiatively efficient systems (i.e. the so-called quasar-mode). Observations of ionized gas winds in AGN have historically been very common (\cite{elvis2000} and \cite{fabian2012}). Most of these past observations, based on long-slit spectroscopy,\footnote{Long-slip spectroscopy employs a narrow slit placed across a celestial object to produce its spectrum in a given spectral band. In general, it probes only a fraction of the source along the direction of the slit, depending on the slit width and seeing conditions.} were unable to assess the spatial scales over which winds occur, preventing us from inferring their mass loss rates,  the impact they exert on to the surrounding ISM/CGM, in particular whether winds regulate star-formation and BH growth. In the last few years, these limitations have been largely overcome thanks to the spectro-imaging capabilities of  integral field units (IFU) on 8/10~m class telescopes, and interferometric observations in the radio-to-(sub)millimetre domain mainly with ALMA, NOEMA and VLA. Interferometry in the radio-to-sub-mm domain, and IFS\footnote{Integral Field Spectroscopy, also called 3D spectroscopy, combines imaging and spectroscopy, probing two spatial dimensions and one spectral dimension. Each pixel in the spatial dimension has its associated spectrum, or spaxel. The main advantage compared to slit spectroscopy is that, according to the source size in the sky and the field of view of the instrument, it allows to observe the whole celestial object.} in the optical/near-infrared, enable to reconstruct with high fidelity the 3D spatial distribution of the gas, the velocity fields and velocity dispersion, allowing to map all kinematic components, including winds.  These capabilities allow a fundamental step in quantifying wind parameters, that is measuring their spatial extension, how far they reach through the host galaxy and in the galactic environment, the CGM, and their global morphology. 

\begin{figure}
\centering
\includegraphics[scale=0.4]{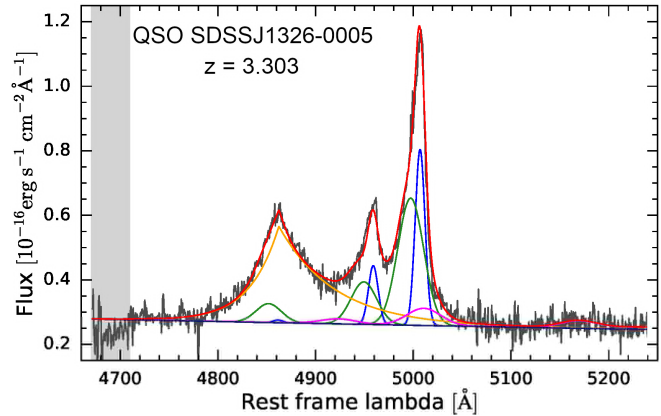}
\caption{The spectrum of a z$\sim3$ QSO in the region of $H\beta$ and [OIII]. The observed profile of the [OIII] doublet (black line) appears skewed toward the blue and shows a broad component. The observed spectrum is modelled by  a combination of narrow components (blue line), a broad component for H$\beta$ arising from the BLR (orange line), and a broad blueshifted component for [OIII] (green line). The latter traces the ionized wind. The magenta line models the FeII emission.  Figure adapted from \cite{bischetti2017}.}
\label{figoiii}
\end{figure}

When using emission lines as probes, for both the warm ionized and the cold molecular phases, winds appear as broad and blue/redshifted emission components,  that, at least in the most luminous QSOs, show high-velocity wings reaching up to $\pm 1000$~km~s$^{-1}$ with respect to the systemic velocity (Figure~\ref{figoiii}). The presence of very high velocity wings in the emission line profiles with emission that is spatially resolved (not point like) is a clear signature of outflows, as, owing to the large velocity, it cannot be ascribed to disc rotation nor inflows. In fact, disc rotation velocity reaches at most few hundred km~s$^{-1}$ in the most massive systems, while inflows are expected in the range few tens km~s$^{-1}$ to $\sim100$~km~s$^{-1}$ \cite{combes2014,Bournaud_et_al_2011}.

However, outflows are not always so fast and extreme, in some AGN their LOS velocity is of the order a few 100~km~s$^{-1}$, so their spectroscopic signatures are hidden into the line profiles determined by the disc rotation. In these cases, dynamical modelling further helps in identifying the ordered kinematics (i.e. rotating discs) and pinpointing deviations from the disc rotation. Any deviations may be identified with either winds, or inflows, or other gas motions related to mergers and tidal events. Based on these techniques, several massive outflows of ionized, neutral and molecular gas, extended on kiloparsec scales, have been discovered in the last fews years, and are today routinely detected in most AGN, both locally and in the high-redshift universe. 

\subsubsection{Warm ionized galactic winds}

Regarding warm ionized gaseous winds, the most used probes are [OIII] $\lambda$ 5007~{\AA} and H$\alpha$ $\lambda$ 6563~{\AA} emission lines,  that trace warm ionized gas in both the host galaxy and in the Narrow Line Region (NLR, see Chapter 1, this Volume), and can be mapped with high resolution both in nearby AGN and up to $z\sim2$--3 with current optical/near-infrared IFUs. At $z\sim2$, a spatial resolution of ~kpc can be achieved by using adaptive optics (AO) -assisted observations. Warm ionized winds are seen in many, perhaps all, local Seyfert galaxies, on scales of the NLR. There, they often, but not always, show a wide opening-angle bi-conical morphology, their speed can reach $\sim$1000~km~s$^{-1}$, and their outflow rates are usually modest, few M$_\odot$~yr$^{-1}$.  Warm ionized winds are common also in luminous QSOs at redshift 2--3, where [OIII] winds reach similar velocities of a few 1000~km~s$^{-1}$, but their mass and outflow rates are much larger compared to local lower luminosity AGN, reaching BH masses of $10^9$--$10^{10}$~M$_\odot$ and M$_\odot$~yr$^{-1}$ of hundreds or even thousands (\cite{bischetti2017}). In high-redshift QSOs, the spatial distribution of these winds is more difficult to assess due to observational limitations and requires AO-assisted observations.  It has been shown that their size can reach several kpc. Both the high velocities ($>$1000~km~s$^{-1}$) of the outflows and their masses suggest that they are driven by the AGN. Some high redshift QSOs show a spatial anti-correlation between the [OIII] wind component and the $H\alpha$-emitting component, that traces on-going star formation in the host galaxy (\cite{canodiaz2012}, \cite{cresci2015}).  This has been interpreted as the direct evidence of QSO driven winds quenching star formation locally within the discs. JWST will enable the  mapping of warm ionized outflows in the first QSOs at the Epoch of Reionization through [OIII] detection. Ionized gas winds are seen also through UV emission and absorption lines. [CIV] $\lambda$ 1549~{\AA} and MgII $\lambda$ 2798~{\AA} emission lines often show extremely asymmetric blueshifted profiles and/or broad components in QSOs, tracing winds that probably arise from the BLR (Broad Line Region, \cite{vietri2018}). Even if they carry a significant amount of energy, they can hardly reach the scales of the host galaxy. Strong outflows are often studied in the rest-frame UV also via blueshifted broad absorption lines \cite{weymann91}. These features are typically characterised by velocity of $v\lesssim15000$~km~s$^{-1}$, but relativistic BALs with $v\sim0.1$--$0.3c$ have been also found in several luminous QSOs \cite{bruni2019, bischetti2022}. The spatial extent associated with BALs outflows detected in luminous QSOs  may reach few hundred pc but is badly constrained. Their kinetic power is estimated to be high enough to play an important role in the co-evolution of SMBHs and their host galaxies
\cite{fiore2017}.

\subsubsection{Cold neutral and molecular winds}

The investigation of winds in the neutral-atomic gas phase winds relies mostly on the Na I D $\lambda$ 5890, 5896~{\AA}, the far-infrared emission lines from atomic carbon [CI] $^3P_1 - ^3P_0$ ($^3P_2 - ^3P_1$) at 492.160 (809.341) GHz, and [CII] $^2P_{3/2} - ^2P_{1/2}$ at 1900.537~GHz rest frame, and the  H I 21 cm line (1.420 GHz).  The Na I D (ionization potential of 5.1~eV) winds require shielding by dust and therefore are most often found in nearby ultra-luminous infrared galaxies (ULIRGs) hosting AGN, where they typically reach regions of few kpc in size (\cite{rupke2005}). The [CII] 1900.537~GHz emission line has been used to directly measure the outflow size in high redshift QSO (\cite{maiolino2012}), whereas the [CI] emission line has also been proposed to be promising wind tracers in nearby sources (\cite{krips2016}). Fast H I outflows, detected through 21 cm line blue-shifted absorption against the radio continuum, are found in particular in young radio sources, and can be driven both by the radio jet and also by wide-angle nuclear winds (\cite{morganti2018}). In recent years, thanks to the gain in bandwidth and sensitivity of radio telescopes, several AGN-driven outflows of atomic hydrogen  have been discovered in the spectra of active galaxies. They are traced by very broad, blueshifted absorption wings with velocities ($>1000$~km~s$^{-1}$) that are much larger than typical rotation velocities of galactic discs. Differently from emission line winds, H I winds are only detected as blueshifted absorption lines and do not show any redshifted component (inverse P-Cygni profile), i.e. indicating that the gas seen against the radio continuum is outflowing. Because their size and location are usually loosely constrained, the neutral gas mass and outflow rates are rather uncertain, and probably reach a few tens  M$_\odot$~yr$^{-1}$. Often the HI outflows do have molecular counterpart.  

Finally, there are winds that involve the cold molecular  ISM phase. These are particularly interesting for feedback processes because they involve directly the star-formation reservoir.  Molecular winds are observed through both emission and absorption lines, similarly to ionized gas neutral gas winds. The most used tracers are bright emission lines such as the CO, HCN, HCO$^+$  rotational transitions that occur in the rest-frame sub-mm, and absorption lines such as OH that occur in the far-infrared. The latter have been mainly detected with Herschel as P-Cygni or blueshifted profiles in nearby ULIRGs hosting AGN (\cite{sturm2011}).  

As ionized gas winds, the molecular ones present a  broad range of LOS velocities (from few 100 to $>1000$~km~s$^{-1}$), and the velocity is usually correlated with the AGN luminosity, the fastest winds residing in the most luminous QSOs. As for the regions reached by molecular winds, because the absorption line are detected against the radio-far-infrared continuum, the determination of the outflow size is limited by the instrumental resolution, but evidences are that OH outflows occur on scales of few 100 pc (\cite{gonzalez2017}). Molecular outflows seen in emission, mainly through CO, HCN, on average reach sizes of $\sim$ kpc, i.e. the typical sizes of the galactic bulge in massive galaxies. There are also some cases where molecular outflows reach much larger scales in the CGM, as in the case of radio sources showing powerful radio jets (\cite{nesvadba2009}).

\begin{figure}[h]
\centering
\includegraphics[width=\textwidth]{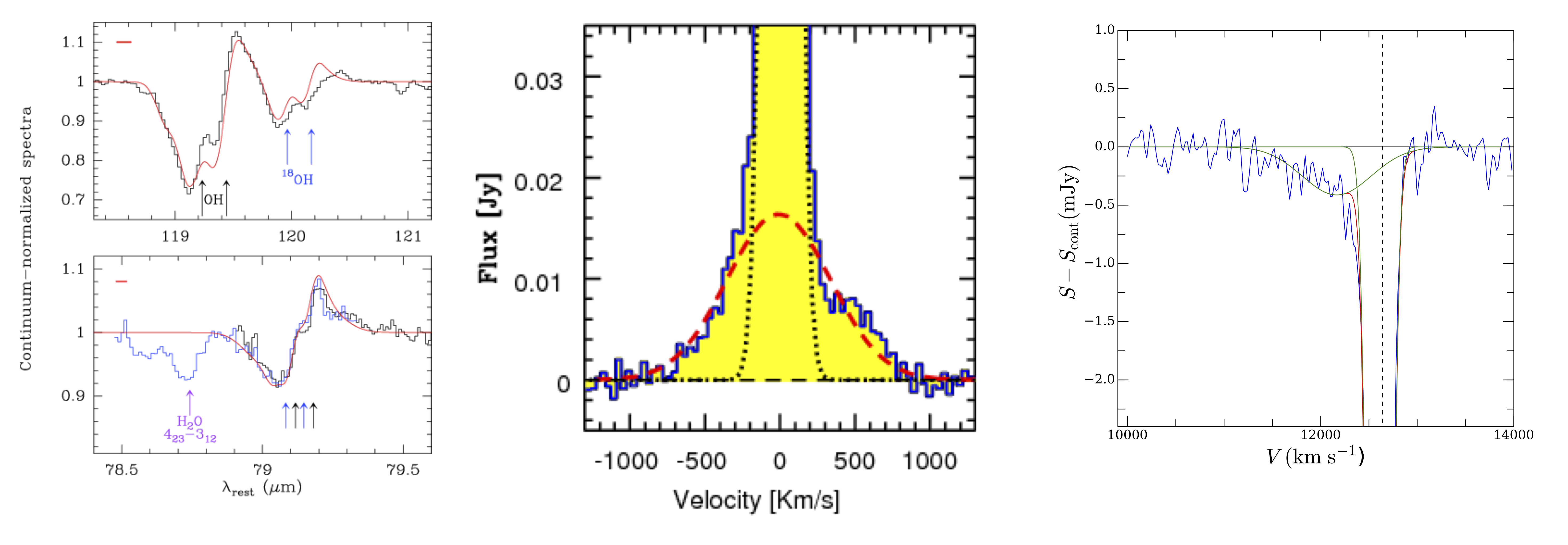}
\caption{An example of a cold wind detected in multiple tracers in the same source, the local ULIRG/AGN Mrk 231. The molecular wind is seen as blue-shifted and P-Cygni profiles of far-infrared  OH transitions (left-hand panel, from \cite{fischer2010}) and as broad wings in the CO(1-0) line profile (central panel, \cite{feruglio2010}). The cold atomic wind is traced by the broad blueshifted wing of the H I 21 cm absorption line (right-hand panel, \cite{morganti2016}).}
\label{figmol}
\end{figure}

In some cases, like the local ULIRG/AGN Mrk231, it was possible to detect the cold wind both in absorption, through Herschel PACS OH lines in the far-infrared and H I 21 cm line in the radio, and through multiple molecular emission lines (CO, HCN, HCO$^+$, Figure~\ref{figmol}). In Mrk231, the OH absorption coupled with radiative transfer models indicate that OH arises from a more compact region compared to the region mapped by the CO, HCN outflows, which, instead occur on kpc scale and are co-spatial (\cite{feruglio2015,aalto2012,gonzalez2017}). The origin of the cold gas in outflows is still a matter of debate.  One possibility is that the cold gas forms in situ as a result of a fast cooling sequence in the post shock region (\cite{richings2018}). Evidences are that HCN, HCO$^+$ may be enhanced in winds compared to the CO emission, as a consequence of shock chemistry. According to \cite{1989ApJ...340..869N}, HCN or HCO+ enhancement occurs because strong UV radiation produced at the shock front can photo-dissociate CO, making C available for further reaction networks leading to HCN and HCO$^+$ formation. HCN shock enhancement was found in shocked regions around young stellar objects (\cite{Tafalla2010}), in the molecular wind of Mrk231 (\cite{aalto2012}) and along the radio jet of M51 (\cite{2015ApJ...799...26M}). An alternative, perhaps less likely scenario involves pre-existing cold gas clouds being accelerated to high speeds in the shock.

The molecular and ionized winds occurring on galactic scales in AGN host galaxies  show outflow mass rates that range from a few to $10^3~\rm M_{\odot}$~yr$^{-1}$ , and wind loading factors, $\eta=\dot M/{\rm SFR}$ systematically higher than that of starburst-driven winds. Both the outflow rate $\dot M$ and the kinetic wind power $\dot E_{\rm kin}$ show correlations with the AGN bolometric luminosity (\cite{fiore2017}). Overall, evidences are that AGN-driven winds might statistically provide the needed wind mass-loading factor to regulate the formation of massive galaxies, $\log M_*/{\rm M}_{\odot}>10.5$, while they are likely less effective on the bulk of the galaxy population with smaller stellar masses.

\subsubsection{Extended X-ray emission and cavities: ISM}%Chiara

Another form of AGN feedback that needs evaluating is that traced by  cavities, or {\it lacunae} in the ISM by the effect of AGN radiative field and/or shocks. Thanks to sensitive spatially resolved X-ray observations, mainly with the Chandra Observatory, coupled with multiband observations, evidences have been reported of extended X-ray emission on scales of tens to hundred pc in nearby Seyfert galaxies. Chandra high resolution imaging of the soft, line-dominated X-ray emission shows remarkable correspondence between high surface brightness X-ray features, the radio jet and optical line emission (\cite{fabbiano2018}), and a spatial anticorrelation with emission of cold molecular gas, giving rise to the so-called molecular cavities. Evidences are reported of cavities on scales of hundreds pc in the cold molecular ISM, as traced by low-J CO transitions,\footnote{Observations of molecular gas in galaxies vastly rely on emission from CO lines arising from states with rotational quantum numbers J $>1$. The CO rotational ladder, or Spectral Line Energy Distribution (SLED), describes the excitation of CO, typically normalized to the ground state. As the rotational levels are increasingly populated, the line intensity increases and the SLED rises until the levels approach local thermodynamic equilibrium (LTE). At LTE the excitation at those levels saturates.} where CO emission is possibly suppressed  due to strong AGN irradiation and/or shocks, that may excite CO up in the rotational ladder, and that are filled with X-ray emitting gas (\cite{rosario2019,shimizu2019,feruglio2020}). The detection of warm molecular gas traced by the H2 2.12 $\mu$m emission line in these CO-cavities supports this scenario, confirming that the scattering material giving rise to the hard X-ray extended emission is likely warm molecular gas.

\subsection{Feedback models}%Chiara & Francesco

\begin{figure}[h]
\centering
\includegraphics[scale=0.7]{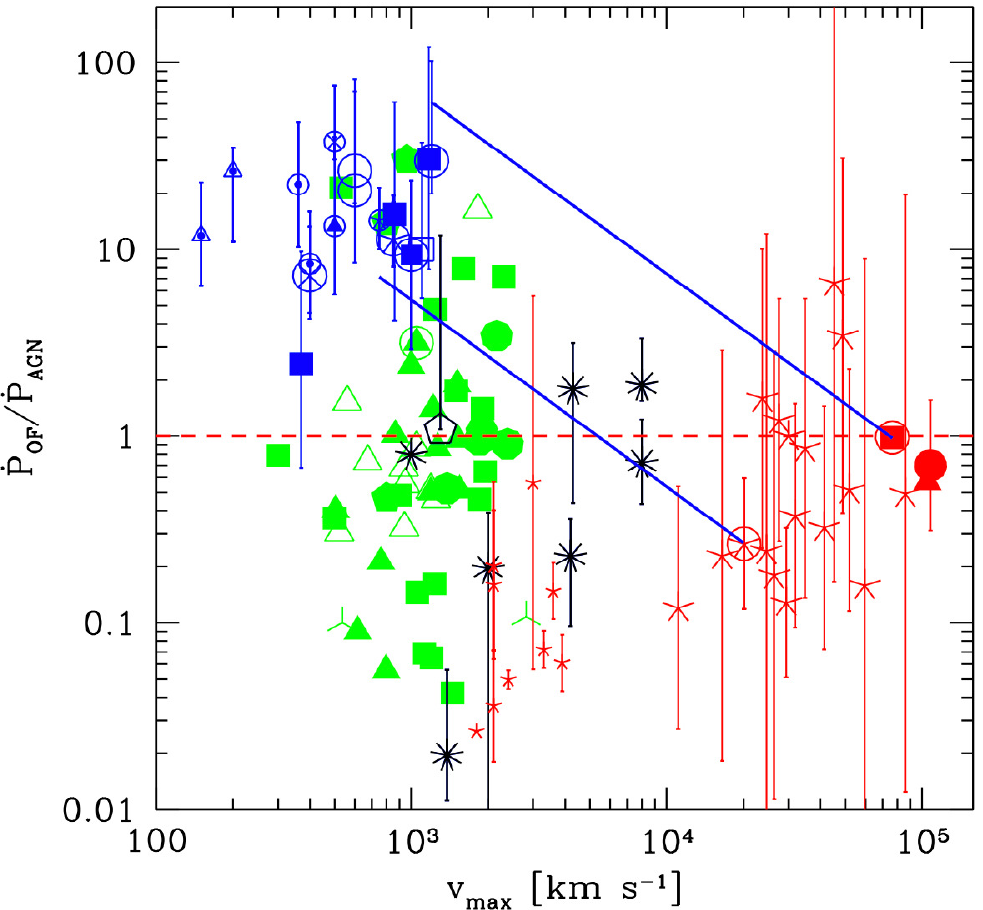}
\caption{Wind momentum load (outflow momentum rate divided by the AGN radiation momentum rate $L_{\rm AGN}/c$) as a function of the outflow velocity $v_{\rm max}$. The red dashed line marks the expectations for a momentum conserving outflow. AGN for which molecular winds have been reported in the literature (mostly local ULIRGs and Seyfert galaxies) are shown with blue symbols. Green symbols mark ionized outflows measurements detected in the optical. BAL winds are shown with black stars. Red symbols mark X-ray UFOs, and the small five point stars are WAs. The two blue solid lines mark the expectations for pure energy conserving outflows for IRAS~F11119$+$13257 and Mrk~231 (figure from \cite{fiore2017}).}
\label{fig:fiore2017}
\end{figure}

SMBH fuelling and feedback processes are often considered as disjoint and studied independently at different scales, both in observations and simulations, but a multi-scale approach is required to understand this phenomenon, which encompasses micro- (mpc-pc), meso- (pc-kpc), and macro-scales (kpc-Mpc) (\cite{gaspari2020}). Observationally, this translates in an inherent multi-wavelength approach. 

Observations of galaxy-scale molecular, neutral and ionized outflows in quasars and ULIRGs have provided evidence to support the idea that feedback driven by AGN winds may be able to effectively influence the host galaxy environment (e.g. \cite{veilleux2020}). Indeed, galaxy-scale molecular outflows can carry a large amount of kinetic luminosity and a large mass outflow rate, and may be capable of removing materials from the host galaxy and also of regulating the star-formation rate. However, the exact mechanism by which the AGN interacts with the molecular gas in the host galaxy and drives the outflows is still debated. Hot X-ray outflows are very important candidates for such an interaction. 

The propagation of the inner wind to the galaxy scale environment can be physically divided in two main regimes: momentum- or energy-driven (e.g. \cite{zubovas2012,faucher2012}). In the energy-driven regime, conservation of energy gives $\dot{P}_{\rm out} = f (v_{\rm in}/v_{\rm out})\dot{P}_{\rm in}$, where $\dot{P} = \dot{M}v$ is the momentum rate or force (``in'' refers to the inner X-ray wind and ``out'' refers to the galaxy-scale outflow, and $v$ is the wind velocity). The efficiency factor $f$ is the fraction of the kinetic power of the inner X-ray wind that goes into bulk motion of the swept-up interstellar material. For a purely energy-driven case, the value would be unity. Instead, in the case of a momentum-driven outflow, most of the energy would be dissipated before accelerating the large-scale outflow and we would expect the relation to be $\dot{P}_{\rm out} \simeq \dot{P}_{\rm in} \simeq L_{\rm AGN}/c$.

IRAS~F11119$+$3257 and Mrk~231 were the first examples of cold galaxy-scale outflows that have been found to be driven by an inner X-ray UFO (\cite{tombesi2015,feruglio2015}). Figure~\ref{fig:fiore2017} shows that the increasing number of sources with simultaneous detection of UFOs and multi-wavelength outflows suggest that the data are well bracketed by the two theoretical momentum- and energy-driven regimes and there is a range of efficiency factors for the coupling of the energetics of the nuclear and galaxy-scale winds that likely depend on specific physical conditions in each object (\cite{fiore2017}).

\section{Extragalactic surveys and statistical populations of AGN}%Ryan

As discussed above, BHs and their host galaxies can exhibit remarkably complex interplay through feeding and feedback. This interplay can be probed in depth through observations, but detailed studies are limited to relatively bright sources. To probe the {\em global} connection between AGN and their hosts, we require statistical studies of populations of AGN and galaxies and their evolution over cosmic time. For this we can use blank-sky extragalactic surveys, which provide a view of all detectable galaxies and AGN within a respective volume. These surveys range from very deep, small area fields covering several arcmin$^2$ (e.g. the GOODS/Chandra Deep Fields \citep{dick03,giac01}) to shallower, wide-field surveys covering much or all of the sky (e.g. SDSS \citep{york00sdss}, Swift/BAT), as well as fields of intermediate depth and area (e.g. COSMOS \citep{civa16cosmos}, Bo\"{o}tes \citep{masi20cdwfs}, XMM-XXL \citep{pier16xxl}). To identify and characterize AGN and galaxies requires observations at multiple wavelengths, and many surveys include coverage in the radio, infrared, optical, and X-rays. Together, these observations yield a picture of the AGN and galaxy population from the Local Volume all the way to redshift 6 and beyond, allowing us to trace the evolution of BHs and galaxies over the full scope of cosmic time. 

\subsection{AGN selection through X-ray surveys and characterization of host galaxies}

X-ray surveys have proven particularly powerful in identifying AGN \citep{bran15xray}. Accreting BHs are typically very bright in X-rays relative to stellar processes in galaxies, and X-rays are also relatively insensitive to moderate obscuration by gas up to hydrogen column densities $N_{\rm H}\sim10^{23}$ cm$^{-2}$ \citep{hick18araa}. X-rays therefore  provide a useful probe of the intrinsic luminosity of an AGN, even when the host galaxy dominates the light in the optical and infrared. Thus many studies have started with samples of X-ray selected AGN, typically identifying sources with luminosities above that expected for stellar processes ($L_X>10^{42}$ erg s$^{-1}$ has been a typical threshold). Due to their relatively low background, surveys with Chandra and XMM-Newton can  detect AGN with remarkably small numbers of source X-ray photons (as few as ten counts or less in some surveys; e.g. \citep{masi20cdwfs}). While the X-ray spectra for bright sources can be modelled in detail, estimating X-ray luminosities for faint sources requires assumptions about the shape of the X-ray spectra.  Other complementary techniques can also be used to select AGN that are X-ray weak or obscured; these include optical line ratios indicating hard sources of ionizing radiation from an accretion disc, mid-infrared colours characteristic of AGN-heated dust, or radio measurements showing non-thermal continuum in excess of that expected for star formation (see e.g. \citep{hick09corr, pado17agn} for more on these methods). Once a sample of AGN is identified, optical and near-infrared images can then be used to determine galaxy morphology and stellar mass ($M_\star$), while photometry in the mid- and especially far-infrared allows for measurements of the SFR at wavelengths with limited contamination from the accreting BH (e.g. \citep{mull11agnsed}).

\begin{figure}[t]
\begin{center}
\includegraphics[width=0.8\textwidth]{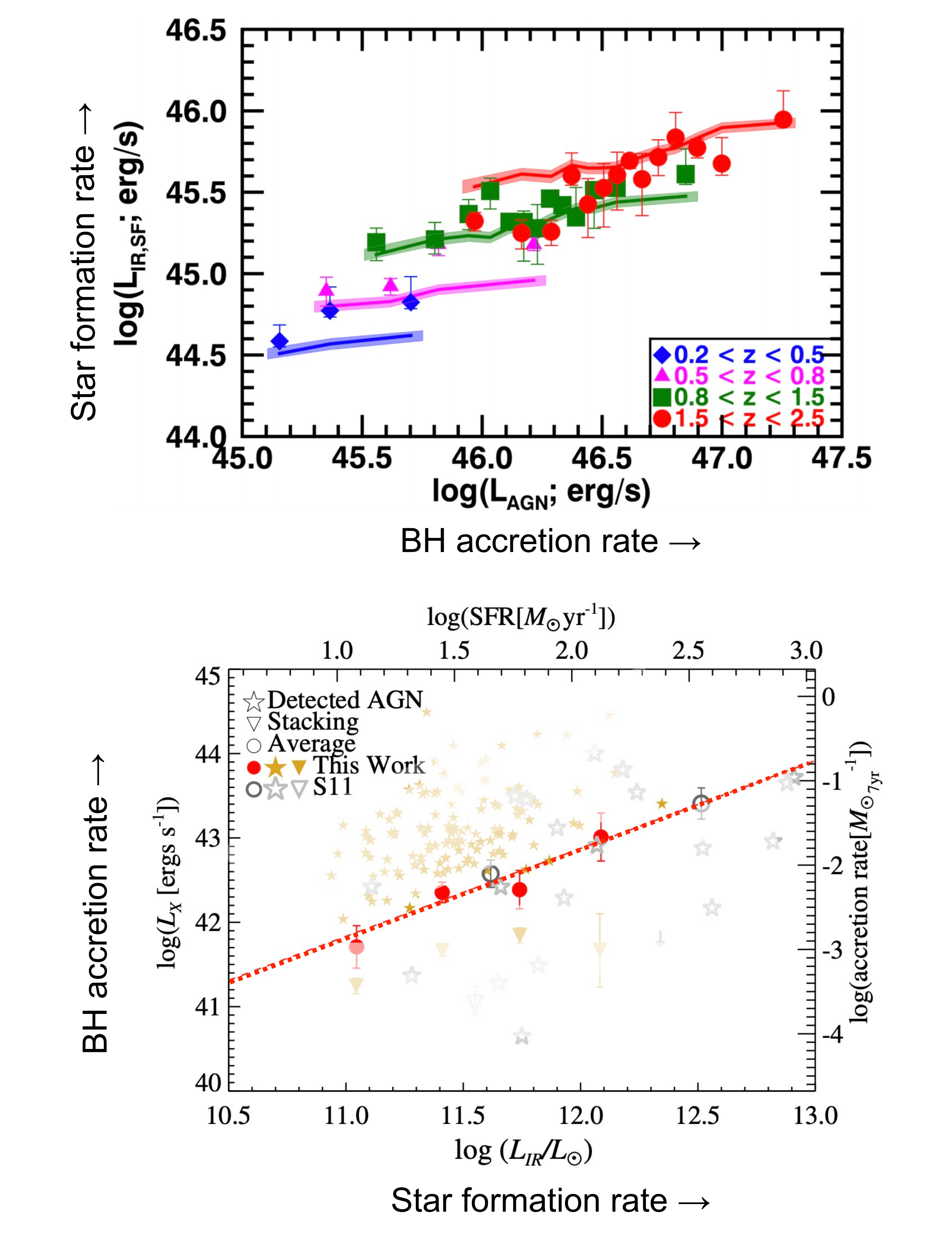}

\caption{Previous studies using far-infrared and X-ray data have yielded little or no connection between BH growth and AGN luminosity ({\em Top}; \citep{stan17qsosf}), or a strong linear correlation ({\em Bottom}; \citep{chen13agnsf}). \label{fig:agnsf}} \end{center}
\end{figure}

\subsection{Connections between BH accretion and star formation}

Using these simultaneous measurements of AGN luminosity (a proxy for BH accretion rate) and SFR, it is possible to probe the associated growth of BHs and their host galaxies. However, studies on this topic have reported somewhat contradictory results. A number of works have found that for X-ray selected AGN, the typical host galaxy SFR has little or no dependence on AGN luminosity, rising with increasing redshift according to the overall cosmic evolution of SFR (see Figure~\ref{fig:agnsf} top panel; e.g. \citep{shao10agnsf, stan17qsosf}). This result points toward there being no causal connection between BH growth and SFR, raising questions about the origin of the BH-galaxy scaling relationships discussed above and the roles of AGN fuelling and feedback. However, other studies have selected all galaxies and averaged the AGN luminosities in bins of SFR, finding a strong, linear connection between BH growth and SFR (Figure~\ref{fig:agnsf} bottom panel; e.g. \citep{chen13agnsf}). This would suggest that BHs and galaxies evolve together, driven by a common supply of gas (but with the precise relationships between them influenced by feedback effects).

To reconcile these seemingly conflicting results, authors have noted the difference in relative time-scales of AGN accretion and SFR. Hydrodynamic BH accretion simulations and observations of ``light echoes'' around recent AGN indicate that AGN ``flicker'' stochastically on time-scales of $10^5$--$10^6$ years or less (e.g. \citep{nova11bhsim, keel12voor, scha15flicker}), suggesting that $L_X$ or other instantaneous AGN luminosity indicators are not good proxies for the long-term BH accretion rate \citep{hick14agnsf}. It is therefore important to measure the distribution in AGN luminosities, and more specifically the Eddington ratio, which gives a measure of the scaled growth rate of the BH. The Eddington ratio requires an estimate of the BH mass; while this can be measured directly for unobscured, broad-line AGN using virial techniques, for most AGN in X-ray surveys the AGN is obscured or the required spectroscopic data are not available. In these cases, Eddington ratio is often approximated by ``specific BH accretion rate'' (sBHAR), determined by the ratio of AGN luminosity and $M_\star$, thus implicitly assuming a relationship between $M_{\rm BH}$ and $M_\star$ (e.g. \citep{aird12agn}).

\begin{figure}[t]
\begin{center}
\includegraphics[width=0.8\textwidth]{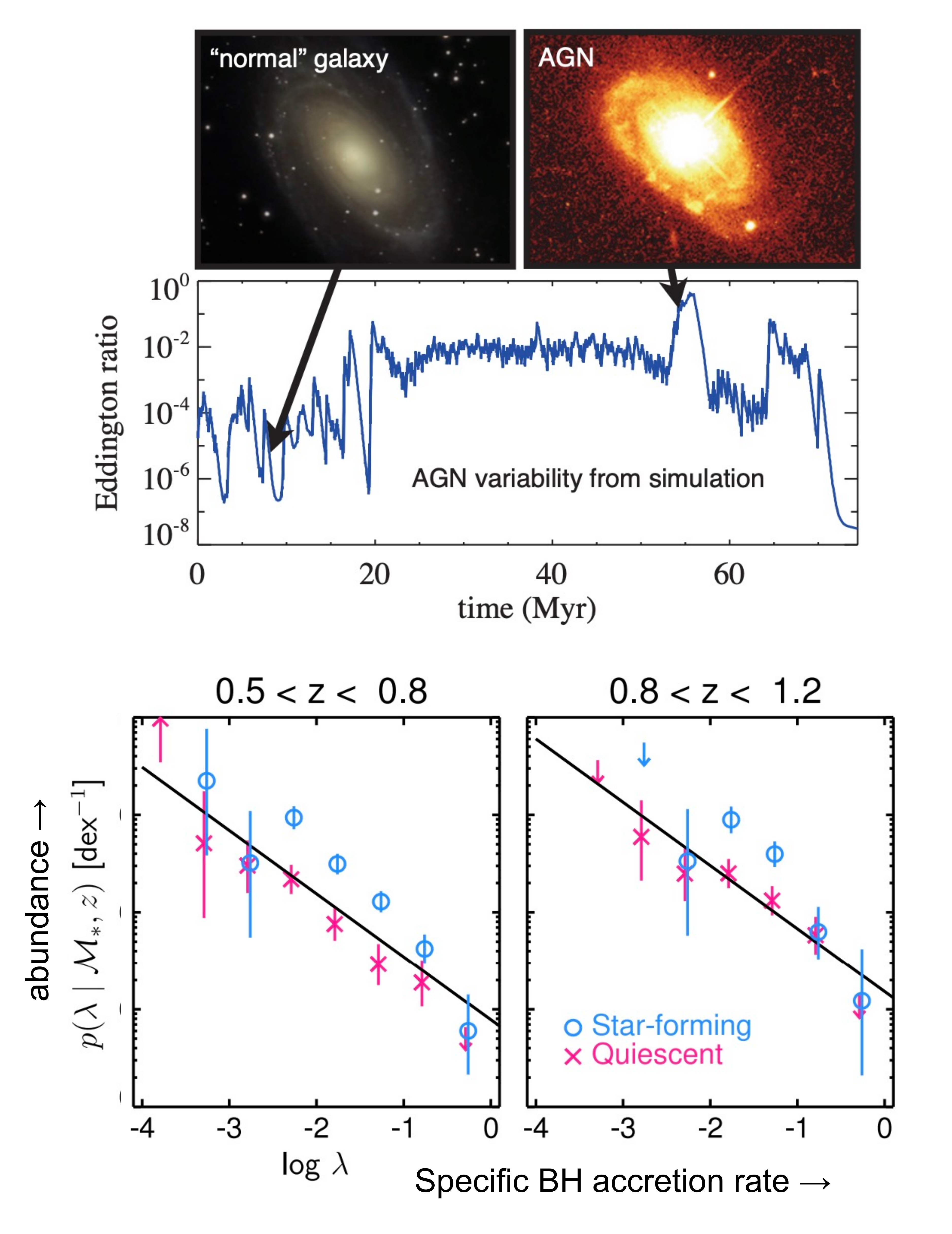}
\caption{{\it Top:} schematic of BH accretion as a stochastic process (including a theoretical light curve \citep{nova11bhsim}), where only the most powerful growth phases tend to be observed as ``AGN'' \citep{hick14agnsf}. {\it Bottom:} the stellar-mass normalized BH accretion rate distributions cover a wide dynamic range, with an approximately power-law distribution that evolves with $z$ and (possibly) SFR \citep{azad15primus}. \label{fig:flicker}} \end{center}
\end{figure}

In X-ray selected AGN studies, the distribution in Eddington ratio (or specific accretion rate) is remarkably similar across different galaxies of different masses (e.g. \citep{aird12agn, bong12xgal}) and with a broad power-law shape that is consistent across various $L_{\rm AGN}$ indicators once flux limits and contamination from star formation are taken into account \citep{jone16sdss}. This wide dynamic range in BH accretion rate also agrees well with the predictions of hydrodynamic simulations that trace both BH feeding and feedback over cosmological time-scales (e.g. \citep{nova11bhsim}). Measurements of the full distribution of sBHAR confirm that the average BH growth evolves with redshift similarly to sSFR \citep[e.g.][]{aird12agn}, and that more highly star-forming galaxies have Eddington ratio distributions shifted to higher BH growth rates (e.g. \citep{azad15primus, aird19agnsf}, Figure~\ref{fig:flicker})

Our understanding of these phenomena has continually grown as data quality has improved, and sample sizes of AGN and galaxy have increased. Today's most sophisticated studies have uncovered interesting trends that challenge the simplest picture of galaxy-BH co-evolution. Detailed measurements of the sBHAR distributions that divide galaxies in terms of $M_\star$ and SFR find a stronger correlation of BHAR with $M_\star$ than with SFR \citep{yang17agngal}. This suggests that the apparent linear relationship between SFR and $L_{\rm AGN}$ may simply be due to the intrinsic correlation between SFR and $M_\star$ for star-forming galaxies at a given redshift \citep{elba11ms}. Therefore galaxies and BHs may not always grow together, as would be suggested by the simplest models in which AGN activity and SF are fuelled by a common supply of gas.

\subsection{Black hole fuelling and galaxy morphologies and mergers}

\begin{figure}[t]
\begin{center}
\includegraphics[width=\textwidth]{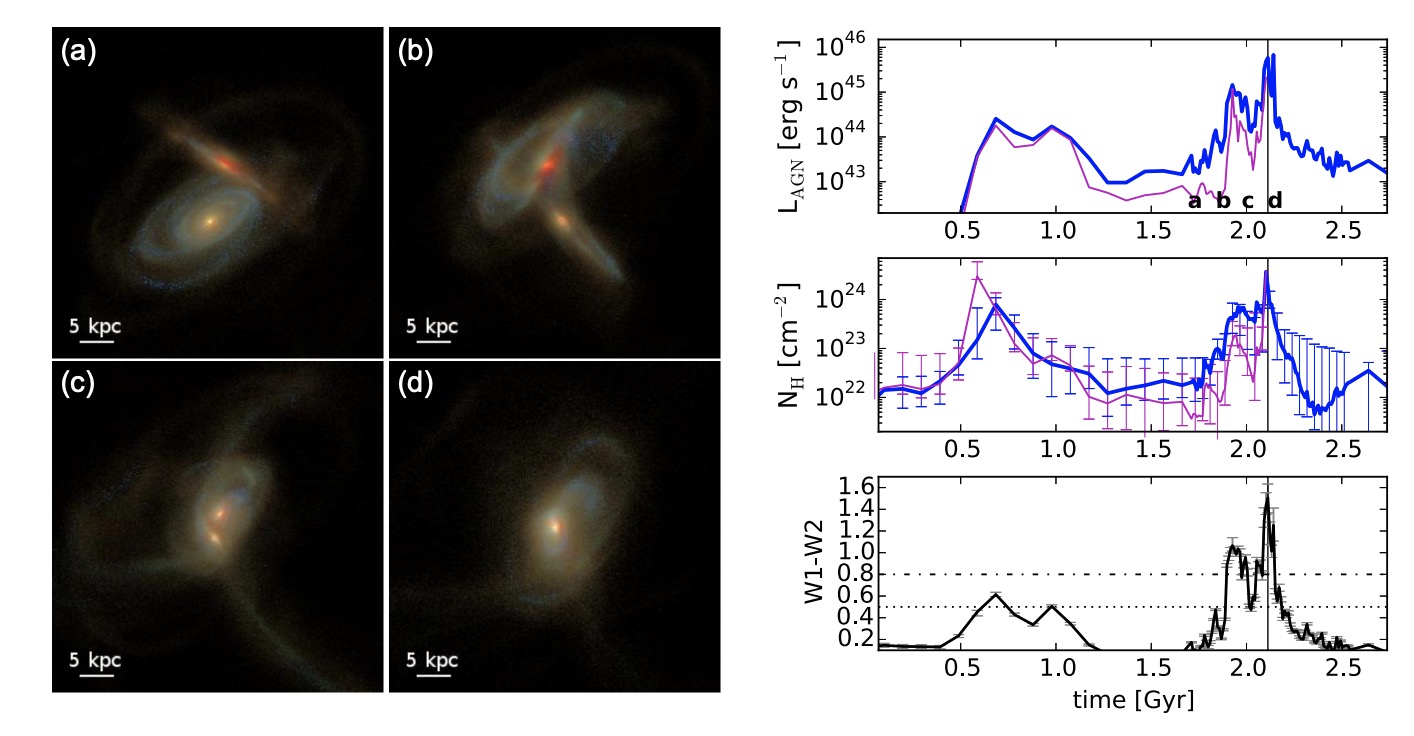}
\caption{Hydrodynamic simulation of AGN fuelling in a galaxy merger, illustrating that the bulk of BH growth ({\em top right}) is at coalescence, with significant obscuration by gas and dust ({\em middle right}). The {\em bottom right} panel shows the WISE $W1$-$W2$ colour during the merger \citep{blec18merger}.
\label{fig:mergers}} \end{center}
\end{figure}

More recent studies, dividing galaxies by $M_\star$, SFR, and galaxy morphology, have indicated that BH accretion is related to SF much more strongly if only  bulge-dominated galaxies are considered \citep{yang19bulge}. Furthermore, the BHAR appears to be mostly closely connected to the central density of stars in the galaxy, consistent with studies that have shown a high fraction of AGN in compact galaxies \citep{koce17compact, ni21compact}. Therefore BH accretion seems to be closely tied to the buildup of dense stellar cores in galaxies, suggesting that the core link is between BHs and the host bulges.

There is compelling observational and theoretical evidence that critically important phases of BH and galaxy evolution occur during galaxy  mergers, which can drive gas to the centres of galaxies that fuel the growth of stellar bulges and accretion on to BHs (e.g. \citep{hopk08frame1}; see Section~1.1). There has been considerable debate as to whether mergers make up a substantial fraction of BH growth, due to the inherent challenges in identifying AGN merger features in survey imaging, as well as issues in selected appropriate control samples of galaxies for which to compare merger rates to AGN hosts. At present the general consensus is that the role of mergers in the overall AGN population is limited (e.g. \citep{cist11agn, mari19merge}). However, it is clear that merging and interacting galaxies are far more likely to host powerful AGN than non-mergers (e.g. \citep{saty14wisepairs, goul18merge, elli19merge}) and can be a key driver of powerful AGN in galaxy formation simulations (e.g. \citep{mcca20merger}). Mergers are also known to drive powerful starbursts (e.g. \citep{sand96}), which are associated with heavy obscuration of the central AGN (e.g. \citep{chen15qsosf, blec18merger}; Figure~\ref{fig:mergers}). However, to date studies have been limited by the number of identified mergers, so it remains unclear what is the precise role of mergers in BH growth, or whether any excess of AGN in mergers can be associated with increased SF and/or central stellar density.

\subsection{Clustering and dark matter halos}

Multiwavelength studies of AGN populations allow us to probe not only the galaxies in which BHs reside, but also their parent dark matter halos and large-scale structures.  Some simulations point to the dark matter halo potential as the key parameter that regulates the ultimate mass of BHs (e.g. \citep{rosa15agnhalo}), and predict that AGN are more likely to reside in central galaxies (rather than satellites) at any given $M_{\rm halo}$, following the trend observed for SF (e.g. \citep{chat12bhsim}). Theoretical studies show that a potential tool for exploring AGN fuelling mechanisms is the halo occupation distribution (HOD), a measure of how AGN or galaxies populate halos. Specifically, the HOD specifically measures, as a function of halo mass, the number of sources that lie at the centres of halos as well as the number that are satellite galaxies in those halos. The observed HOD can potentially distinguish between galaxy interactions, disc instabilities, or accretion from hot halo gas as the dominant trigger for BH growth (e.g. \citep{fani13xclust, gatt16clust}), providing complementary constraints to measurements of host galaxy characteristics.

The dark matter halos of AGN can be determined observationally in multiple ways. One is two-point spatial correlation function, which measures how strongly clustered objects are in space compared to a random distribution, The correlation function of dark matter halos, and its dependence with mass and redshift, can be inferred from numerical simulations, and so these measurements of AGN allow us to associate AGN with the halos in which they reside. Another method uses the weak gravitational lensing produced by the dark matter halos of AGN. This can be measured either via the shapes of background galaxies or distortions of the cosmic microwave background. Early studies of X-ray AGN clustering found that they reside in halos of mass $M_{\rm halo} \sim10^{13}$~M$_\odot$ \citep{hick09corr}, with little variation in redshift and luminosity. This indicated that X-ray selected AGN are found in more and less massive systems than optically-selected broad-line quasars and radio galaxies, for which $M_{\rm halo}$ is a factor of a few smaller and larger, respectively (e.g. \citep{hick09corr}). While there is little dependence on the typical $M_{\rm halo}$ with AGN luminosity, there has been observed a strong dependence on $M_{\rm BH}$ \citep{krum15xclust}, suggesting that the mass of the BH is tied not only to the mass of its host galaxy, but also to the parent dark matter halo.

More recently, studies have moved beyond measurements of a single characteristic $M_{\rm halo}$ to estimates of the full HOD. The first HOD studies of $0.5 < z < 2$ X-ray AGN in the Bo\"{o}tes \citep{star11xclust} and COSMOS \citep{rich13agnhod} fields yielded a strong preference for central galaxies (with a fraction of satellites $f_{\rm sat}<$ a few per cent), with the occupation fraction increasing strongly with $M_{\rm halo}$ as predicted by simulations. In contrast, HOD measurements for optical quasars, which are biased toward higher-Eddington populations than X-ray AGN (e.g. \citep{hick09corr}), obtained a marginally higher $f_{\rm sat}\approx10\pm5$ per cent, and lower average ${M_{\rm halo}}$ compared to X-ray AGN (e.g. \citep{shen13qsohod}). An analysis of the X-ray AGN HOD constrained by weak lensing measurements obtained $f_{\rm sat}$ as high as 18 per cent, although with considerable uncertainties \citep{leat15agnhalo}.  

\subsection{Obscured and elusive AGN}

While most X-ray selected AGN identified in surveys have modest obscuration, studies of the local AGN population and synthesis models of the cosmic X-ray background indicate that a large fraction (as high as 50 per cent) of AGN are heavily obscured column densities up to and above Compton-thickness ($N_{\rm H}>10^{24}$ cm$^{-2}$; \citep{hick18araa, anan19cxb}). For these sources the soft X-ray emission can be heavily suppressed, so that heavily obscured AGN are often not detected even in deep Chandra and XMM surveys. Studying these sources therefore requires modelling the integrated cosmic X-ray background, for which the peak at $\approx$30 keV energies requires a large population of heavily obscured sources (e.g. \citep{anan19cxb}). Alternatively, we can utilize other selection techniques for individual AGN that are less affected by heavy obscuration. Studies of mid-infrared selected AGN have revealed a large population that are weak or undetected in X-rays, but whose average X-ray emission (determined through stacking techniques) is consistent with a high fraction of Compton-thick sources (e.g. \citep{lamb20agn, carr21obsc}).

The physical nature of the obscuring material for the bulk of AGN remains an open question (\citep{hick18araa}; see also Chapter~1). Clearly some AGN are obscured by the a parsec-scale torus \citep{netz15unified}, and recent results suggest that the obscured fraction depends strongly on Eddington ratio, suggesting that the material in the obscuring torus is regulated by radiative feedback from the AGN (e.g. \citep{ricc17edd}). Still, some obscuration in AGN may come from gas on circumnuclear ($\sim$100 pc) or host galaxy ($\gtrsim$kpc) scales. Studies have found that galaxies with higher SFR or disturbed morphologies have a higher fraction of obscured AGN (e.g. \citep{koce15xmerge, chen15qsosf}), and spatial clustering measurements indicate that obscured AGN are more strongly clustered than unobscured AGN (e.g. \citep{dipo17qsoclust, powe18xclust}), although this difference may depend on redshift and luminosity. AGN obscuration therefore cannot be purely explained by random orientation, and may also be associated with a particular phase (such as a galaxy merger) in the evolution of the BH and its host galaxy. 

\section{Prospects for the future and new facilities}

\begin{figure}[h]
\centering
\includegraphics[scale=0.3]{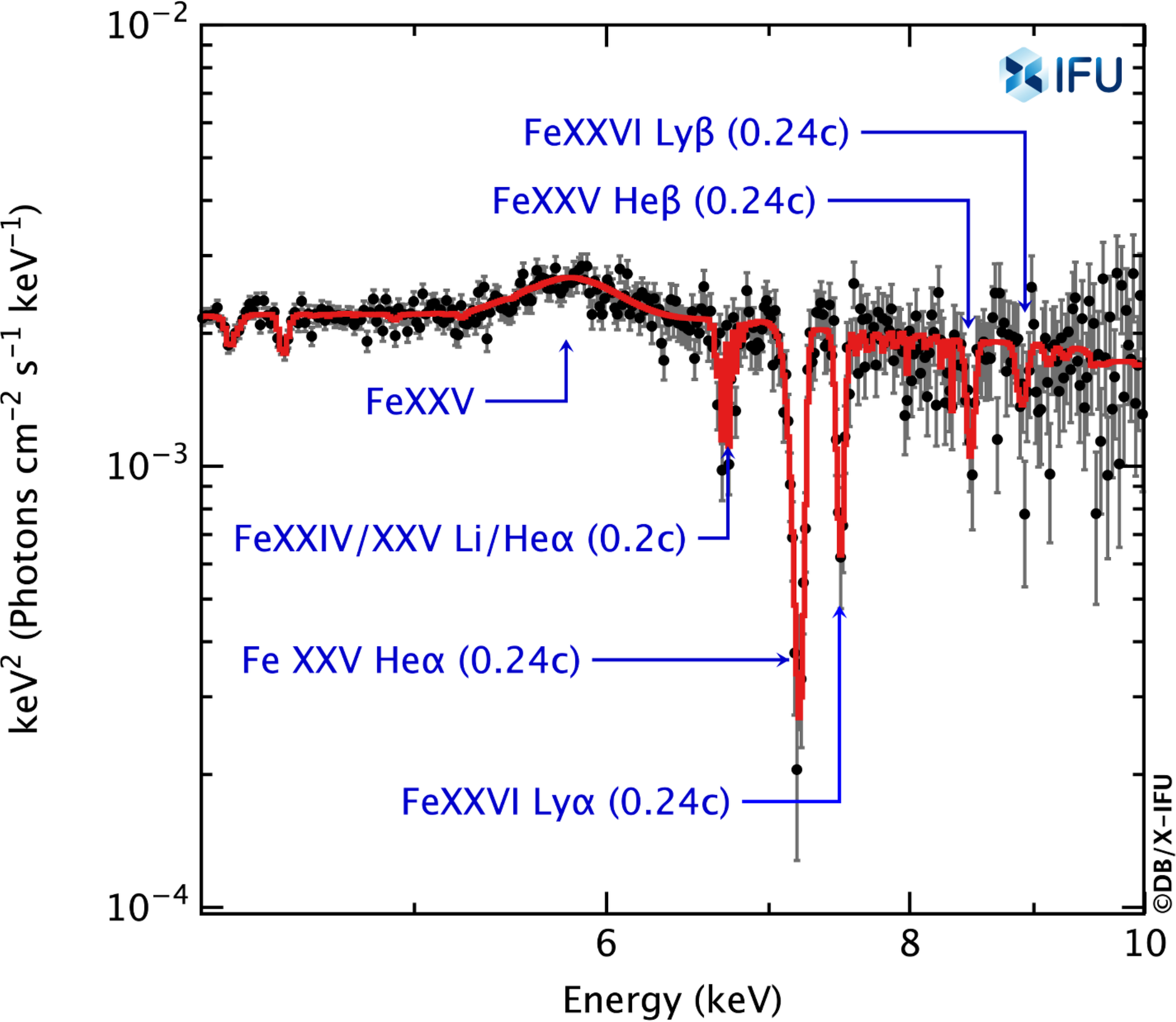}
\caption{Simulated 100 ks \emph{Athena} X-IFU spectrum of the quasar PDS~456. A series of absorption lines from an outflow with two velocity components at $v_{\rm out}=0.20$--$0.24c$ and a turbulent velocity broadening of $3 \times 10^3$~km~s$^{-1}$ would be clearly detectable thanks to the unprecedented high-energy resolution and throughput provided by the \emph{Athena} X-IFU (Credits: X-IFU Consortium).}
\label{fig:x-ifu}
\end{figure}

X-ray observations are a unique way to address the main questions regarding AGN feedback because they probe the initial, highest-ionization and hottest phase of the outflow, which carries most of the kinetic energy (e.g. \cite{tombesi2013}). High throughput, high spectral and spatial resolution instruments as those proposed for the upcoming X-ray Imaging and Spectroscopy Mission (XRISM) (\cite{xrism2020}), and the future \emph{Athena} (\cite{barcons2017}), \emph{Lynx} (\cite{ozel2018}), and \emph{AXIS} (\cite{mush2018}) are required to determine the acceleration and launching mechanism(s) of SMBH driven outflows and to quantify their impact on galaxy evolution.\\

The key to progress on this investigation is a detailed characterization of the physical properties of these winds (column density, ionization state, outflow velocity, location, geometry, covering factor, etc.). \emph{XRISM} will provide high spectral resolution observations but, due to the relatively low collecting area and spatial resolution, these will be limited to the nearby brightest AGN. Only the high-energy resolution and high throughput offered by the proposed \emph{Athena} and \emph{Lynx} will allow the study of X-ray outflows on a large enough sample of sources to effectively probe the prevalence and evolution of these systems from the early universe to the current epoch (e.g. \cite{geor2013}). These proposed missions will provide a high enough signal-to-noise ratio to utilize short time-scale ($\simeq$ hrs) variability as a tool to explore the wind launching region down to a few Schwarzschild radii.

The unprecedented data quality will allow us to seek correlations among the fundamental parameters such as density, ionization, outflow velocity, and luminosity, which will uniquely constrain predictions of radiation-driven (e.g. \cite{proga2000}), momentum-driven (\cite{king10outflow}), and magnetically-driven (\cite{fukumura2010}) accretion disc wind models. Figure~\ref{fig:x-ifu} shows a 100~ks simulated \emph{Athena} X-ray integral field unit (X-IFU) spectrum of the quasar PDS~456, which will allow a full determination of the outflow kinematic and ionization structure, necessary for reliably constraining the wind energetics. Such X-ray observations are key for determining the total column density, the highest velocities, the highest ionization, and hence the kinetic power carried by SMBH outflows, which is required to drive the ensuing galaxy-scale outflows. 

For AGN populations and demographics, surveys with new missions have the potential for dramatic steps forward in depth and statistical power. Recently, the eROSITA mission has surveyed the entire sky to unprecedented depth in soft X-rays, yielding very large samples of X-ray selected AGN (e.g. \citep{brun22efeds}). Looking forward, planned or proposed X-ray observatories (including \emph{Athena} as well as multiple concepts being developed for potential NASA X-ray missions) would provide deep X-ray imaging over large areas. In combination with sensitive multiwavelength surveys to characterize the redshifts and host galaxies of these AGN, these missions extend the ability to study AGN demographics to the first few billion years of cosmic time, and even place constraints on the formation of the first galaxies and BHs (\cite{civa19bh}).

\bibliographystyle{spbasic-FS-PRC}
\bibliography{section12_chap3_capelo,section12_chap3_feruglio,section12_chap3_hickox,section12_chap3_tombesi}

\end{document}